\documentclass[%
 reprint,
superscriptaddress,
 amsmath,amssymb,
prb,
]{revtex4-1}

\usepackage{graphicx}
\usepackage{dcolumn}
\usepackage{bm}
\usepackage{appendix}
\usepackage{soul}
\usepackage[utf8]{inputenc}
\usepackage[english]{babel}

\usepackage[T1]{fontenc}
\usepackage[colorlinks,bookmarks=false,citecolor=blue,linkcolor=red,urlcolor=blue]{hyperref}
\usepackage[dvipsnames]{xcolor}
\usepackage{times}
\usepackage{bbm}
\usepackage[normalem]{ulem}
\usepackage{multirow}
\usepackage{booktabs}

\newcommand{\be}{\begin{eqnarray}}
\newcommand{\ee}{\end{eqnarray}}

\begin{document}


\title{Vortex-lattice melting and paramagnetic depairing in the nematic superconductor FeSe}

\author{F.~Hardy}
\affiliation{Institute for Quantum Materials and Technologies, Karlsruhe Institute of Technology, 76021 Karlsruhe, Germany}

\author{L.~Doussoulin}
\affiliation{Univ. Grenoble Alpes, CNRS, Grenoble INP, Institut N\'eel, F-38000 Grenoble, France}
\affiliation{Institute for Quantum Materials and Technologies, Karlsruhe Institute of Technology, 76021 Karlsruhe, Germany}

\author{T.~Klein}
\affiliation{Univ. Grenoble Alpes, CNRS, Grenoble INP, Institut N\'eel, F-38000 Grenoble, France}

\author{M.~He}
\affiliation{Chongqing Key Laboratory of Soft Condensed Matter Physics and Smart Materials, College of Physics, Chongqing University, Chongqing 401331, People's Republic of China}

\author{A.~Demuer}
\affiliation{Univ. Grenoble Alpes, INSA Toulouse, Univ. Toulouse Paul Sabatier, EMFL, CNRS, LNCMI, 38000 Grenoble, France}

\author{R.~Willa}
\affiliation{Institute for Theoretical Condensed Matter Physics, Karlsruhe Institute of Technology, 76131 Karlsruhe, Germany}

\author{K.~Willa}
\affiliation{Institute for Quantum Materials and Technologies, Karlsruhe Institute of Technology, 76021 Karlsruhe, Germany}

\author{A.-A.~Haghighirad}
\affiliation{Institute for Quantum Materials and Technologies, Karlsruhe Institute of Technology, 76021 Karlsruhe, Germany}

\author{T.~Wolf}
\affiliation{Institute for Quantum Materials and Technologies, Karlsruhe Institute of Technology, 76021 Karlsruhe, Germany}

\author{M.~Merz}
\affiliation{Institute for Quantum Materials and Technologies, Karlsruhe Institute of Technology, 76021 Karlsruhe, Germany}

\author{C.~Meingast}
\affiliation{Institute for Quantum Materials and Technologies, Karlsruhe Institute of Technology, 76021 Karlsruhe, Germany}

\author{C.~Marcenat}
\affiliation{Univ. Grenoble Alpes, CEA, IRIG, PHELIQS, LATEQS, F-38000 Grenoble, France}

\date{\today}
 
\begin{abstract}
The full $H-T$ phase diagram in the nematic superconductor FeSe is mapped out using specific-heat and thermal-expansion measurements down to 0.7 K and up to 30 T for both field directions. A clear thermodynamic signal of an underlying vortex-melting transition is found in both datasets and could be followed down to low temperatures. The existence of significant Gaussian thermal superconducting fluctuations is demonstrated by a scaling analysis, which also yields the mean-field upper critical field $H_{c2}$(T). For both field orientations, $H_{c2}$(T) shows Pauli-limiting behavior.  Whereas the temperature dependence of the vortex-melting line is well described by the model of Houghton {\it et al.}, Phys. Rev. B {\bf 40}, 6763 (1989) down to the lowest temperatures for H $\perp$ FeSe layers, the vortex-melting line exhibits an unusual behavior for fields parallel to the planes, where the Pauli limitation is much stronger.  Here, the vortex-melting anomaly is only observed down to T$^{*}$ $\approx 2-3$ K, and then merges with the $H_{c2}$(T) line as predicted by Adachi and Ikeda, Phys. Rev. B {\bf 68} 184510 (2003). Below T$^{*}$, H$_{c2}$(T) also exhibits a slight upturn possibly related to the occurence of a Fulde-Ferrell-Larkin-Ovchinnikov (FFLO) state. 

\end{abstract}

\maketitle

\section{\label{Intro}Introduction}

In 1957 Abrikosov~\cite{Abrikosov57} predicted that a magnetic field can penetrate a superconductor as an array of vortices, each carrying a magnetic flux quantum $\Phi_0=h/2e$. This occurs in type-II superconductors in which the normal-superconducting surface energy  is negative, {\it i.e.} when the Ginzburg-Landau parameter $\kappa=\lambda/\xi$, the ratio of the London penetration depth $\lambda$ to the coherence length $\xi$, exceeds the threshold value $1/\sqrt{2}$.~\cite{deGennes} Vortices repel each other and  typically crystallize into a hexagonal lattice. In the presence of weak and randomly distributed disorder, {\it e.g.} point defects,~\cite{Larkin70,Larkin74,Larkin79} this long-range periodicity is lost and a new (dislocation-free) state of matter, still displaying well defined diffraction peaks - the so-called 'Bragg glass' - is formed~\cite{Giamarchi94,Giamarchi97}. 

 In increasing magnetic field, the density of vortices increases until they overlap at the upper critical field $H_{c2}$(T) where superconductivity disappears at a second-order phase transition~\cite{deGennes,Sarma}. However, thermally induced and/or static disorder can lead to a melting of the vortex solid well below the upper critical field $H_{c2}$(T) (for reviews see Refs ~\onlinecite{Blatter94,Brandt95,SchriefferBook}).
 
 This possibility has been first considered by Eilenberger~\cite{Eilenberger67}, but attempts to observe it in low-dimension geometries, to enhance thermal fluctuations, remained unsuccessful~\cite{Huberman79,Fisher80}. Unequivocal thermodynamic evidence of a genuine vortex-lattice melting transition finally came out soon after the discovery of the cuprate superconductors, in which thermal fluctuations are greatly enhanced due to their high $T_c$, very short coherence length and large anisotropy. In a very limited number of exceptionally high-quality single crystals, vortex melting manifests as a tiny discontinuity in the reversible magnetization~\cite{Zeldov95,Welp96,Liang96,Nishizaki96}, while specific-heat~\cite{Schilling96,Schilling97,Schilling98,Roulin95,Roulin96,Roulin98,Revaz98,Bouquet01} and thermal-expansion~\cite{Lortz03M} measurements exhibit a peak superimposed on a step indicative of the additional degrees of freedom in the high-temperature vortex-liquid phase. This first-order transition represents the only genuine phase transition for superconductivity in a magnetic field, since $H_{c2}(T)$ becomes a broad crossover. Since then calorimetric features related to melting were also reported for conventional low-$T_c$ superconductors {\it e.g.} Nb$_3$Sn, SnMo$_6$S$_8$ and for Fe-based superconductors~\cite{Lortz06,Petrovic09,Mak13,Koshelev19}. However, the fate of the melting transition for $T\rightarrow 0$ remains unclear since the low-temperature/high-field region is usually inaccessible, as in cuprates due to the high values of H$_{c2}$(0), or because residual disorder disrupts the melting transition~\cite{Bouquet01}.  Theoretically, the melting line for T $\rightarrow$ 0 may i) be suppressed due to quantum fluctuations~\cite{Blatter93,Blatter94-1,Blatter94-2}, or ii) merge with H$_{c2}$(T) or iii) even disappear at finite temperature in strongly Pauli-limited superconductors~\cite{Adachi03}.

The Zeeman effect represents another mechanism which can affect the high-field superconducting phase transitions, and intense research efforts in low-$T_c$ unconventional superconductors including organics, ruthenates and heavy fermion, have focused on the emergence of high-field phases where this effect is dominant (also referred to as strongly Pauli-limited superconductors). Prominent examples are  $\kappa$-(BEDT-TTF)$_2$Cu(NCS)$_2$~\cite{Lortz07fflo} and CeCoIn$_5$~\cite{Bianchi03,Matsuda07}, which exhibit thermodynamic evidence of a modulated phase having Cooper pairs with nonzero total momentum and a spatially non-uniform order parameter $\Delta(\bf r)$~\cite{Fulde64,Larkin65}. While for the former the high-field phase appears to be a physical realization of the original Fulde-Ferrell-Larkin-Ovchinnikov state (FFLO)~\cite{Lortz07,Wosnitza18}, the modulated phase in CeCoIn$_5$ is believed to result from a particular coupling between {\it d}-wave superconductivity and a field-induced incommensurate spin-density wave (SDW)~\cite{Kenzelmann08,Kenzelmann10,Kumagai11}.

The recently discovered Fe-based superconductors offer another interesting platform for the study of vortex matter. As anticipated theoretically~\cite{Murray10}, thermal fluctuations of intermediate magnitude between cuprates and conventional materials, accompanied by a clear vortex-melting anomaly, were highlighted in the 122 and 1144 families via high-resolution thermodynamic measurements~\cite{Mak13,Hou15,Koshelev19}. In parallel, a first-order superconducting transition detected in the magnetostriction of KFe$_2$As$_2$ stressed the relevance of Zeeman depairing in these materials and raised the possibility of observing a FFLO phase~\cite{Burger13,Zocco13,Cho17}.  

Among the Fe-based materials, FeSe has attracted considerable interest as superconductivity emerges deep inside a non-magnetic but electronic nematic phase that breaks 4-fold rotational symmetry below $T_s$ = 90 K~\cite{Boehmer17,Coldea18}. Superconductivity is argued to arise from a spin-nematic pairing driven by orbital-selective spin fluctuations~\cite{Sprau17,Rhodes18,Kreisel17,Kang18,Benfatto18,Cercellier19}. Despite its low $T_c \approx 9$ K, FeSe can be considered as a high-T$_c$ superconductor because of its very low carrier density\cite{Terashima14,Yang17} and Kasahara {\it et al.}~\cite{Kasahara14} have argued that it lies deep inside the Bardeen-Cooper-Schrieffer/Bose-Einstein-Condensate (BCS/BEC) crossover. In this context, the same authors have claimed i) that a field-induced phase transition of the Fermi liquid with strong spin imbalance occurs for $H\parallel c$ within the superconducting state at a field $H^*$ at which the Zeeman energy becomes comparable to $\epsilon_F$~\cite{Kasahara14,Watashige17} and ii) that a genuine FFLO phase is observed for $H\perp c$~\cite{Kasahara19}. Thus, both thermal fluctuations and paramagnetic effects are expected to be large in this high-$\kappa$ superconductor. The moderate value of $H_{c2}(0)$ < 30 T offers a unique opportunity to study the $H-T$ phase diagram down to the lowest temperatures in clean single crystals. 

In this Article, using thermodynamic probes on high-quality FeSe single crystals, we demonstrate the existence of sizable field-induced Gaussian superconducting fluctuations using a scaling approach and provide compelling thermodynamic evidence of the existence of an underlying vortex-melting transition. Our analysis of these data also clearly reveals that Pauli depairing exerts a large influence on the vortex-melting properties in high magnetic fields, in particular for $H \parallel$ FeSe layers. Here we find that the vortex-liquid phase disappears for T $\lesssim$ 3 K, {\it i.e.} below which the vortex-melting line merges with the $H_{c2}$(T) line. Interestingly, such a merging is predicted by mean-field theory~\cite{Adachi03}. Here, it occurs near the expected tricritical point from which the FFLO phase could emerge in very clean single crystals~\cite{Brison97}. Finally, our results exclude that FeSe lies within the BCS/BEC crossover, and we find no thermodynamic signature of the reported high-field phase for $H \perp$ FeSe layers~\cite{Kasahara14,Watashige17,Kasahara19}.  

This Article is organized as follows. In Sec.~\ref{Methods}, the experimental methods (crystal growth and specific-heat and thermal-expansion measurements) are explained in detail. In Sec.~\ref{Results}, we present our raw specific-heat and thermal-expansion data,  which already provide clear evidence for the existence of both large superconducting fluctuations and a vortex-melting transition. Scaling analysis of our thermodynamic data is presented in Sec.~\ref{Discussion} and the resulting $H-T$ phase diagram is analyzed thoroughly using existing models of the mixed state. The possible occurence of the FFLO state is discussed and a consistent check of our analysis is provided. Conclusions are given in Sec.~\ref{Conclusion}.
\section{\label{Methods}Experimental methods}
\begin{figure}[t]
\centering
\begin{minipage}{1\columnwidth}
\centering
\includegraphics[width=1.0\columnwidth]{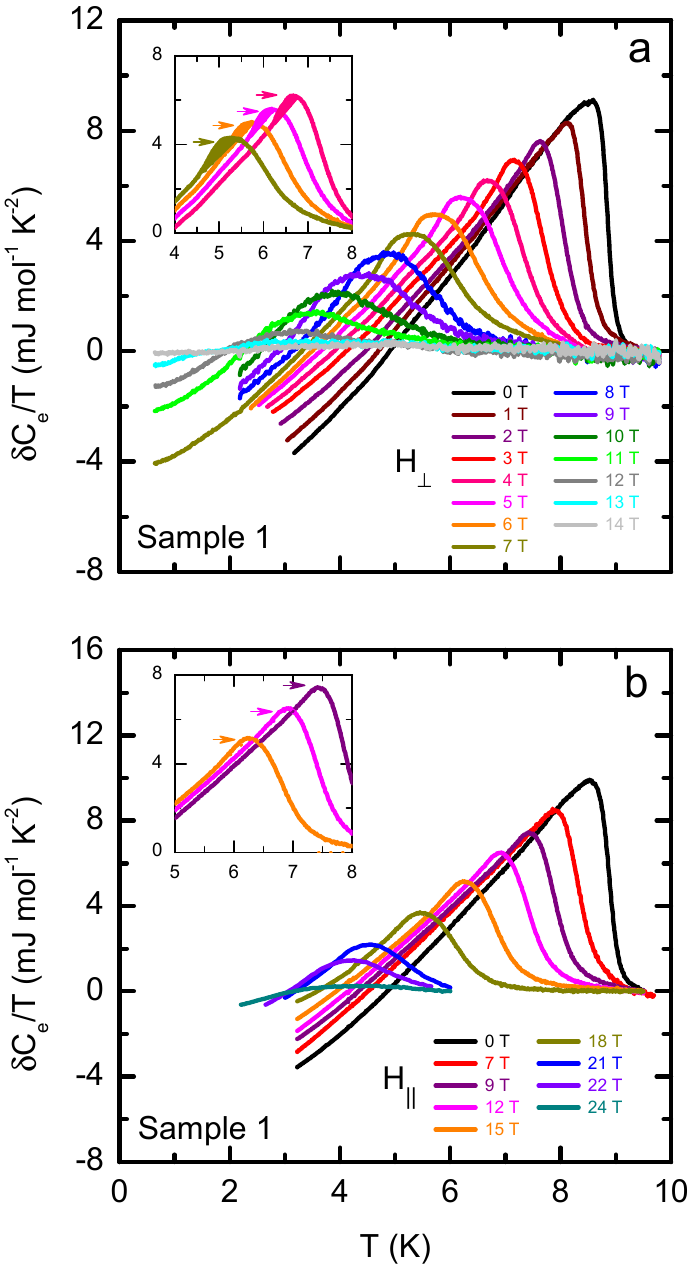}
\end{minipage}
\caption{(Color online) (a)-(b) Temperature dependence of $\delta C_e$/T, the difference between the superconducting- and normal-state specific heats of Sample 1 for H $\perp$ and $\parallel$ to the FeSe layers, respectively. The insets highlight the excess specific heat (shaded area) ascribed to the vortex melting transition T$_m$(H) (see arrows).} 
\label{Fig1}
\end{figure}

Stoichiometric single crystals of FeSe were synthesized by chemical vapor transport using a eutectic mixture of KCl and AlCl$_{3}$ and characterized using single-crystal x-ray diffraction. Samples 1 and 2, used in this work, were taken from batches 2 and 5 of Ref.~\onlinecite{Hardy19}, respectively. 

Specific-heat measurements were performed on Sample 1 up to 30 T and down to 0.6 K using a home-made miniature AC calorimeter. It consists of a bare Cernox chip from Lakeshore Cryotronics Inc. split in two parts and suspended to a small copper ring with PtW($7~\%$) wires. One part is used as an electrical heater while the other part is employed to record the temperature oscillations (a few \% of the average sample temperature in the range 1-10 Hz , see Ref.~\onlinecite{Michon19} for further details). A precise \textit{in-situ} calibration and corrections for the thermometer magnetoresistance were accounted for in the data analysis. This setup allows to measure the specific heat of minute samples with an accuracy better than $\approx 5~\%$, inferred from measurements on 6N copper, with a signal/noise ratio of about $10^4$. High-resolution thermal-expansion measurements were carried out on Sample 2 using a home-built capacitance dilatometer~\cite{Meingast90} with a typical relative resolution $\Delta L/L$ $\approx$ 10$^{-8}$-10$^{-10}$ in fields up to 10 T. 

\section{\label{Results}Experimental results}

Figs~\ref{Fig1}(a)-(b) display the temperature dependence of $\delta C_e(T,H)=C_s(T,H)-C_n(T,H)$, the difference between the superconducting- $C_s(T,H)$ and the nematic-state $C_n(T,H)$ specific heats, for magnetic fields applied perpendicular ($\perp$) and parallel ($\parallel$) to the FeSe layers, respectively. Here, $C_n(T,H)=\gamma_n T + B_3T^3$ was determined by fitting the 18 T-data where superconductivity is fully suppressed down to 0.5 K. The inferred values of the Sommerfeld coefficient and the Debye term amount to $\gamma_n$ = 6.5 mJ mol$^{-1}$ K$^{-2}$ and $B_3$ = 0.4 mJ mol$^{-1}$ K$^{-4}$, respectively, in good agreement with previous reports~\cite{Hardy19,Sun17}. Figs~\ref{Fig2}(a)-(b) display the corresponding in-plane thermal expansion $\delta \alpha_e(T,H)=\alpha_s(T,H)-\alpha_n(T,H)$ measured upon heating (solid line) after cooling in a magnetic field (dashed line). Here, the normal-state contribution $\alpha_n(T,H)$ was determined by fitting the 10 T-data for $T \geq$ 10 K.

A well-defined discontinuity is observed in the zero-field specific heat at T${_c}$ = 8.9 K, with a width of about 1 K related to disorder (twin boundaries and/or a very small number of Fe vacancies~\cite{Watashige15,Jiao17}) indicating the transition from the nematic to the superconducting state. A similar anomaly of comparable width, but with a slightly higher  $T{_c}$ = 9.1 K, is found in our thermal-expansion data. We note that the anomaly is very mean-field-like, in the weak coupling limit, {\it i.e.} at odds with the $\lambda$ transition expected in the 3d-XY universality class for an interacting Bose-Einstein condensate~\cite{Schneider}, which superfluid $^{4}$He belongs to. This rules out that FeSe actually lies deep within the BCS/BEC crossover for which a cusp-like anomaly is expected. For an exhaustive discussion of the BCS/BEC crossover in a two-band model, we refer to Ref.~\onlinecite{Chubukov16}, and in the context of high-temperature superconductivity to Ref.~\onlinecite{Chen05,Schneider}.

\subsection{\label{Evidence_Fluctuation}Large superconducting fluctuations}
The strength of thermal fluctuations is usually  quantified by the Ginzburg number~\cite{Blatter94,Brandt95} given by 
\begin{equation}\label{Eq1}
Gi=\frac{1}{2}\left(\frac{1}{4\pi\mu_0}\frac{k_B T_c \tilde{\Gamma}}{\tilde{H}^2_{c}(0) \tilde{\xi}^3_{\perp}(0)}\right)^2,
\end{equation}
where $\tilde{\xi}_{\perp}(0)$, $\tilde{H}_{c}(0)$ and $\tilde{\Gamma}=\frac{\tilde{\xi}_{\perp}(0)}{\tilde{\xi}_{\parallel}(0)}$ are the respective Ginzburg-Landau values of the in-plane coherence length, thermodynamic critical field and $H_{c2}$ anisotropy expressed in SI units. These can all be inferred from our thermodynamic data. Here $\tilde{H}_{c}(0)$ = 0.21 T and $\tilde{\xi}_{\perp}(0)$ = 3.5 nm are calculated from the zero-field specific heat and the initial slope of $H_{c2}^{\perp}(T)$ (see Sec.~\ref{Hperp}), respectively. $\tilde{\Gamma}\approx4.5$ is inferred from Fig.~\ref{Fig5} where we find that the specific-heat curve for $H_{\perp}=1-2$ T matches that of $H_{\parallel}=7$ T. We obtain $Gi=5\times10^{-4}$ for FeSe, which is several orders of magnitude larger than in classical superconductors, {\it e.g.} Nb ($Gi\approx$ 10$^{-11}$), but slightly lower than in cuprate superconductors (10$^{-1}$ < $Gi$ < 10$^{-3}$)~\cite{Koshelev19}. This large value of $Gi$ strongly suggests that thermal fluctuations cannot be neglected in FeSe.

\begin{figure}[b]
\centering
\begin{minipage}{1\columnwidth}
\centering
\includegraphics[width=1.0\columnwidth]{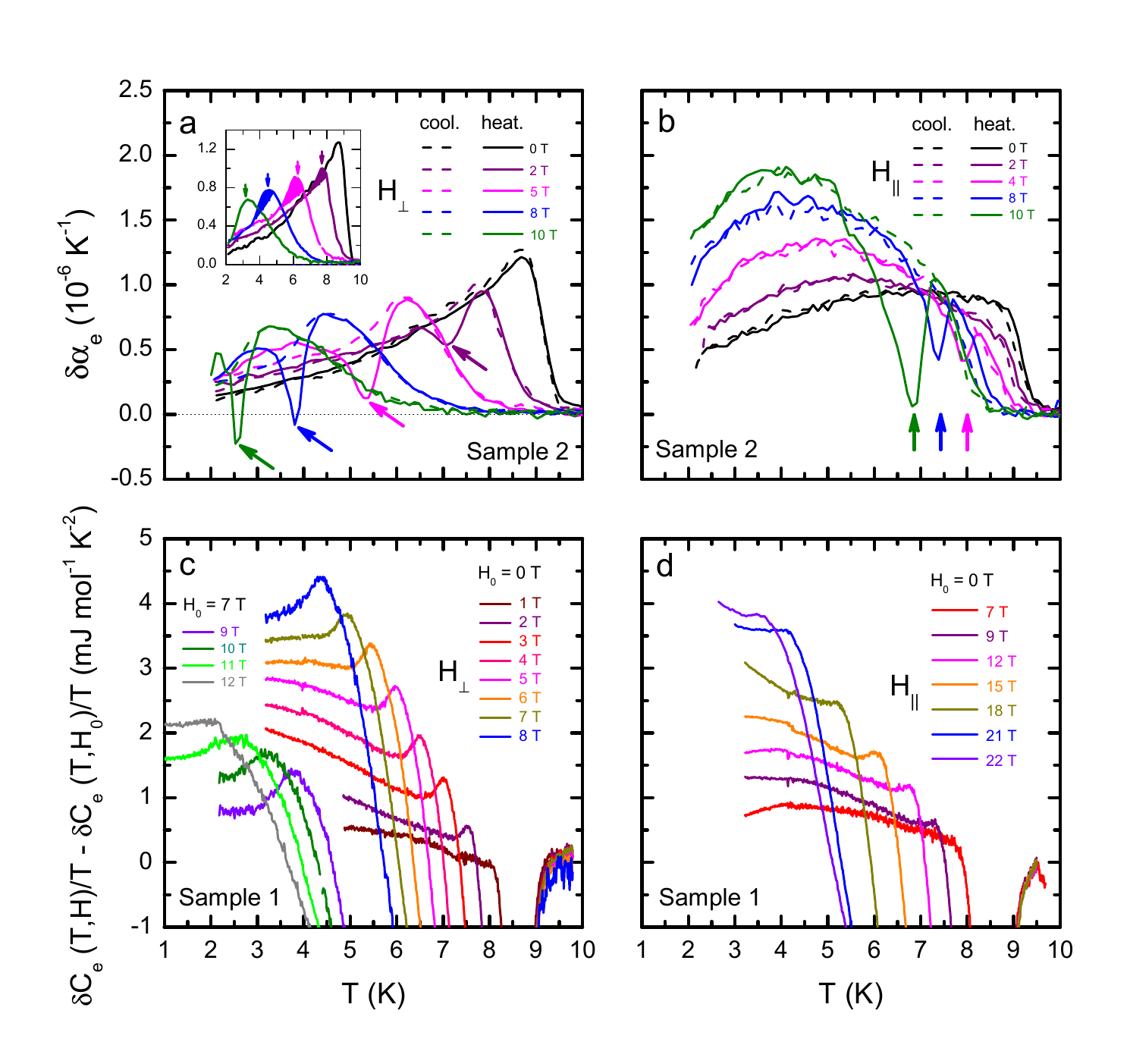}
\end{minipage}
\caption{(Color online) (a)-(b) Temperature dependence of $\delta \alpha_e$, the difference between the superconducting- and normal-state thermal expansions of Sample 2 for H $\perp$ and $\parallel$ to the FeSe layers, respectively. Measurements were carried out upon heating (solid line) after cooling in a magnetic field (dashed line). The inset highlights the excess thermal expansion (shaded area) related to the vortex-melting transition T$_m$(H) (see arrows). (c)-(d) Specific heat of Sample 1 as a function of temperature for H $\perp$ and $\parallel$ to the FeSe layers, respectively. To emphasize the discontinuity at T$_m$(H), the 0 T-data (7 T-data) are subtracted from those obtained for H$_{\perp} \leq$ 8 T (for H$_{\perp} >$ 8 T). In (b), only the 0 T-data are subtracted.} 
\label{Fig2}
\end{figure}

\begin{figure}[t]
\centering
\begin{minipage}{1\columnwidth}
\centering
\includegraphics[width=1\columnwidth]{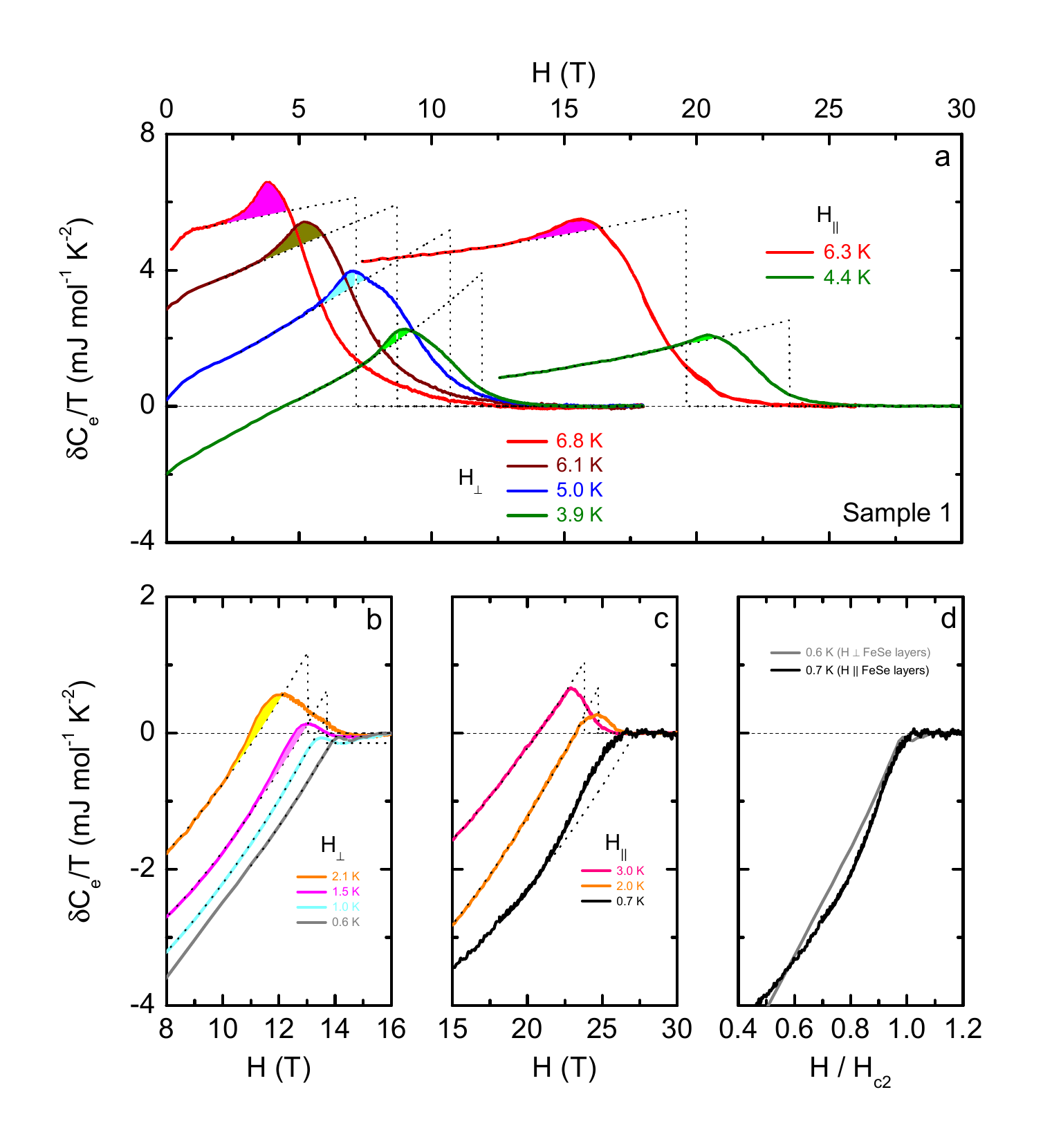}
\end{minipage}
\caption{(Color online) (a) Field dependence of $\delta C_e(H,T)/T$, the difference between the mixed- and the normal-state specific heats of Sample 1 for both $H$ $\parallel$ and $\perp$ FeSe layers at the indicated temperatures. The shaded areas indicate the excess specific heat related to the vortex-melting transition at H$_m$(T). (b)-(c) same as in (a) but at lower temperatures. No melting anomaly is observed for $H\parallel$ layers (c) and only a broadened mean field jump, vanishing below $\approx 1$ K, is observed. (d) Comparison of $\delta C_e(H,T)/T$ for the two orientations, plotted as a function of $H/H_{c2}$, at T = 0.6-0.7 K.}
\label{Fig3}
\end{figure}

A telltale sign of thermal fluctuations is a broadening of the superconducting transition in magnetic field~\cite{Farrant75,Lortz03,Lortz06}. As shown in Figs~\ref{Fig1} and~\ref{Fig2}(a)-(b), a significant broadening of the superconducting transition for both field directions in both measurements is observed. This becomes particularly evident for $H_{\perp} \geq$ 2 T and $H_{\parallel} \geq$ 7 T, where the broadening clearly exceeds the intrinsic transition width. Evidence for large fluctuations are also quite prominent in the field-sweep measurements displayed in Fig~\ref{Fig3}(a), {\it e.g.} for T $<$ 6.8 K where the transition to the normal state extends over several Teslas for both field orientations. A quantitative analysis of this broadening is obtained by the scaling analysis presented in Sec.~\ref{Scaling}. As expected for thermally induced fluctuations, this broadening finally reduces progressively with decreasing temperatures for $T <$ 3 K (see Figs.~\ref{Fig3}(b)-(c)). 

\subsection{\label{Evidence_Melting} Evidence for an underlying vortex-melting transition}
Beside this large broadening of the transition, a small anomaly is clearly resolved on the upward side of the heat capacity anomaly, particularly for $H_{\perp} \geq$ 2 T and $H_{\parallel} \geq$ 12 T (see shaded area in the insets of Figs~\ref{Fig1} and ~\ref{Fig2}(a)). To make this feature more visible and to facilitate comparison with other superconductors, we have subtracted the $H_{\perp}$ = 0 T (resp. $H_{\perp}$ = 7 T) data from those obtained for $H_{\perp} \leq$ 8 T (resp. $H_{\perp} >$ 8 T), as illustrated in Figs~\ref{Fig2}(c)-(d). The broad remaining discontinuity is very reminiscent of the vortex melting transition $T_m$(H) initially reported by Roulin {\it et al.}~\cite{Roulin96} on twinned YBa$_2$Cu$_3$O$_{6.94}$ single crystals and more recently in Ba$_{0.5}$K$_{0.5}$Fe$_{2}$As$_{2}$~\cite{Mak13} and RbEuFe$_{4}$As$_{4}$~\cite{Koshelev19}. It was interpreted as a second-order melting transition between a vortex glass and a vortex liquid~\cite{Roulin98-2}.  A clear signature of vortex melting is also visible in the $H$-sweep measurements where specific heat in excess (shaded areas in Fig.~\ref{Fig3}) is detected for both field orientations.  Interestingly, this melting anomaly, which represents the only genuine phase transition in a hard type II superconductor with strong fluctuations, is clearly less pronounced for $H$ $\parallel$ FeSe layers. Indeed it is only observed in the range 7 $< H_{\parallel}$ $<$ 21 T (see Figs ~\ref{Fig2}(d) and ~\ref{Fig3}(c) and no melting anomaly could be detected below $\approx 3$ K (see Fig.~\ref{Fig3}(c)) where only a broadened mean-field feature persists which in turn vanishes for $T \leq 0.7$ K.

Additional evidence for the existence of an underlying vortex-melting transition is obtained from thermal-expansion measurements, which have proven to be a very sensitive and complimentary probe of the vortex matter~\cite{Lortz03M,Mak13}. As shown in the heating curves (after field-cooling) of Figs~\ref{Fig2}(a)-(b), extra peaks that grow in magnitude with increasing field, are detected slightly below the broadened superconducting transition, at positions which coincide rather well with $T_m(H)$ defined as the mid-point of the broadened discontinuity in specific-heat measurements. Their absence in the cooling curves reveals that they are not electronic in origin, but rather related to irreversible magnetostrictive effects at the melting/irreversibility line due to flux pinning. Similar irreversible peaks were already observed at the melting transition of both YBa$_2$Cu$_3$O$_{7 - \delta}$~\cite{Lortz03} and Ba$_{0.5}$K$_{0.5}$Fe$_{2}$As$_{2}$~\cite{Mak13} crystals with weak flux pinning. Lortz {\it et al.}~\cite{Lortz07,Mak13} argued that they are caused by flux gradients that yield non-equilibrium screening currents which form during cooling and suddenly vanish at the melting temperature upon heating. The applied field exerts a force on these currents which is transferred to the crystal lattice via the pinning centers.  These types of anomaly in the cuprates were found to exhibit a behavior comparable to a kinetic glass transition and are most likely related to some glassy vortex phase~\cite{Lortz07,Lortz02}, whose phase line however corresponds rather well to the first-order melting line in fully reversible crystals.

\section{\label{Discussion}Discussion}
\subsection{\label{Scaling}Scaling of the specific heat and thermal expansion}

A natural way to study thermal fluctuations is to examine the predicted scaling behavior of the specific heat. The same scaling relations apply to the reversible thermal expansion which is closely related to the specific heat through the Ehrenfest or Pippard relations~\cite{Pippard}.

\begin{table*}[t]
\begin{center}
\caption{Superconducting parameters of FeSe. $\tilde{\xi}(0)$ and $\tilde{\Gamma}$ are the Ginzburg-Landau values of the coherence length and $H_{c2}(T)$ anisotropy inferred from the initial slope of $H_{c2}$(T) (see Sec.~\ref{Hperp}) and Fig.~\ref{Fig5}, respectively. $\tilde{\kappa}$ is the Ginzburg-Landau parameter and $Gi$ the Ginzburg number derived in Sec.~\ref{Evidence_Fluctuation}. $H_{orb}(0)$, $H_{p}(0)$ and $\alpha_M$ are the orbital and Pauli critical fields and the Maki parameter determined in Sec.~\ref{phase}.}
\label{Table1}
\begin{tabular*}{.9\linewidth}{@{\extracolsep{\fill} } l c c c c c c c @{\qquad}}
\\
 &
  $\tilde{\xi}(0)$ (nm) &
  $\tilde{\Gamma}$ &
  $\tilde{\kappa}$ &
  $Gi$ &
  $H_{\mathrm{orb}}(0)$ (T)&
  $H_{p}(0)$ (T)&
  $\alpha_\mathrm{M}$ \\
  \hline
$H \perp c$ &
  3.5 &
  4.5 &
  94 &
  5 $\times$ 10$^{-4}$ &
  20 &
  26.5 &
  1.0 \\
$H \parallel c$ &
  0.8 &
  4.5 &
  420 &
  5 $\times$ 10$^{-4}$ &
  90 &
  29 &
  4.4
\end{tabular*}
\end{center}
\end{table*}

In optimally-doped YBCO, thermal-expansion measurements have demonstrated the existence of 3d-XY fluctuations over a $\pm$ 10 K interval around $T_c$~\cite{Pasler98,Meingast01} and their persistence in field up to 11 T without phonon-background subtraction~\cite{Lortz03}.For comparison, analysis of calorimetric data were inconclusive mainly because the fluctuation component represents at most 5 \% of the large phonon contribution~\cite{Inderhees91,Schnelle93,Overend94,Breit95,Junod94,Roulin95-2,Pierson95,Pierson96,Pierson98,Jeandupeux96,Junod00}. In FeSe, the phonon subtraction is straightforward because $T_c$ is low and superconductivity is fully suppressed for $H _{\perp}> $16 T. However, an analysis in zero field is impossible since the intrinsic transition width $\approx$ 1 K clearly exceeds the width of the critical region $\left|{T-T_c}\right| \leq Gi T_c\approx$ 4 mK. On the other hand, in this critical XY regime, the field introduces an additional length scale $\ell_H=\sqrt{\phi_0/\pi H}$ which reduces the effective dimensionality~\cite{Lortz03,Schneider} leading to a broadening of the transition. According to Jeandupeux {\it et al.}~\cite{Jeandupeux96}, the magnetic field breaks the XY symmetry if the correlation length $\xi_{XY}\approx \xi_{0}t^{-2\nu}$ exceeds $\ell_H$ ($\nu_{XY}$ = 0.669 is the XY critical exponent and $t=T/T_c-1$). Thus, fluctuations are expected to be observed only for $H \leq$ $H_{XY}=2 H_{c2}'T_c Gi^{2\nu_{XY}}$. Using $H_{c2 \perp}'=\left(\partial H_{c2 \perp}/\partial T\right)_{T_c}$ $\approx$ 3 T K$^{-1}$ in FeSe (see Sec.~\ref{Hperp}) we obtain $H_{XY}\approx$ 2 mT clearly indicating that this regime can be neglected. Thus we restrict our analysis to the 3d Lowest Landau Level (3d-LLL) framework~\cite{Larkin}.
\begin{figure}[t]
\centering
\begin{minipage}{1\columnwidth}
\centering
\includegraphics[width=1\columnwidth]{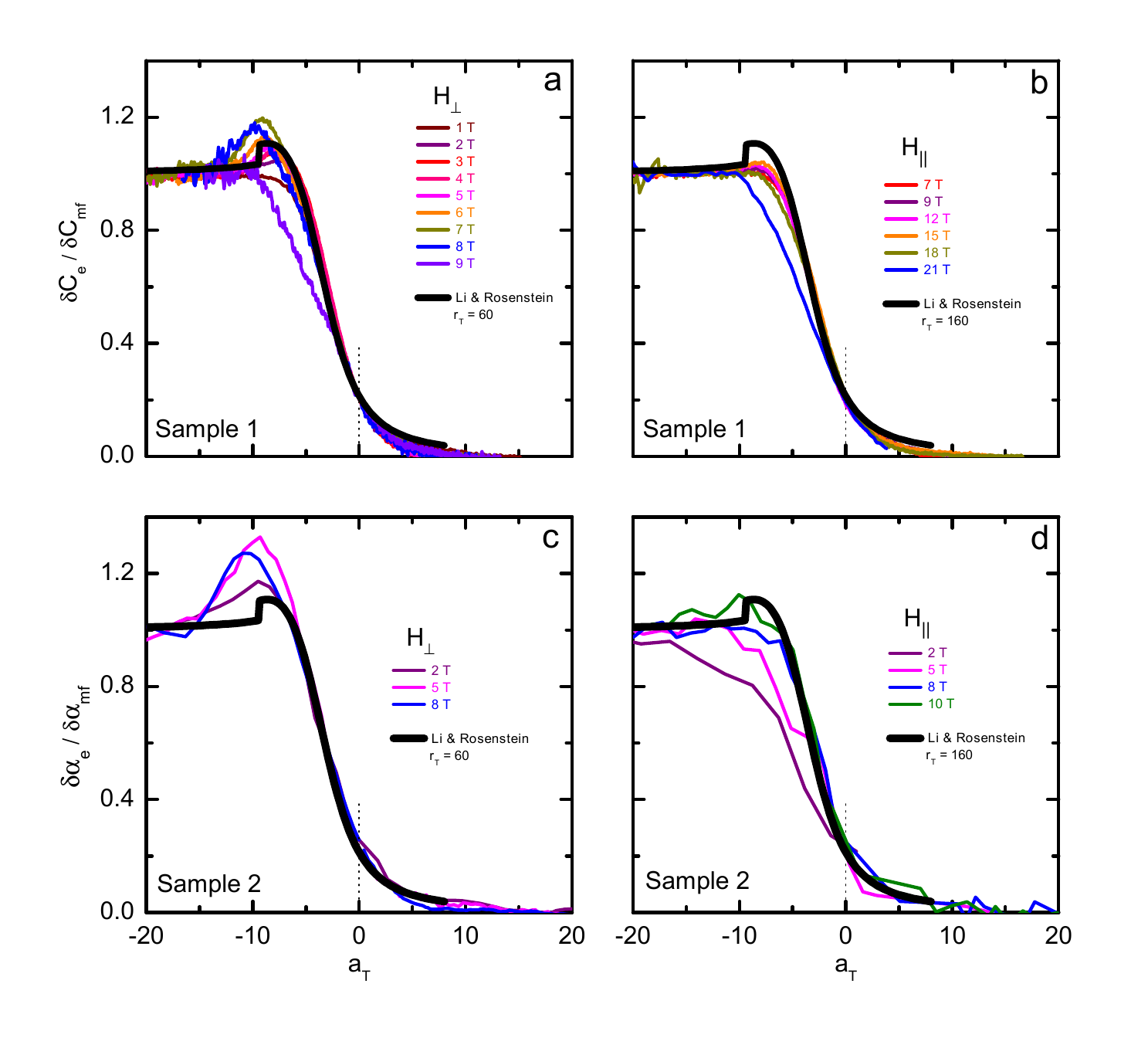}
\end{minipage}
\caption{(Color online) (a)-(b) 3d-LLL scaling of the T-dependent specific heat of Sample 1 for H $\perp$ and H $\parallel$ to the FeSe layers, respectively. (c)-(d) Same for the thermal expansion of Sample 2 (cooling curve). The thick line represents the scaling function calculated by Li and Rosenstein~\cite{Li01,Li01-2,Li03,Li04,Rosenstein10} for the given values of $r_{T}$ (see text for details). The discontinuity at $a_T$ = -9.5 corresponds to the vortex-lattice melting transition. The dashed line indicates the mean-field $T_{c}$(H).}
\label{Fig4}
\end{figure}

In this model, the broadening of the transition is enhanced in field by $Gi(H)=Gi^{1/3}\left(H/H_{c2}'T_c\right)^{2/3}$~\cite{Larkin}. It is applicable if the field is high enough to confine the Cooper pairs in their lowest Landau level, {\it i.e} for $H > H_{LLL} = GiH_{c2}'T_c\approx$ 10 mT for $H \perp$ FeSe layers. In the vicinity of the mean-field transition temperature $T_c(H)$, Thouless~\cite{Thouless75} has shown that $\delta C_{e}(T,H)$ normalized by $\delta C_{mf}(T,H)$, i.e. the difference in heat capacities expected from mean-field theory, is a universal function of the single scaling variable
\begin{equation}\label{Eq4}
a_T=r_T\left[\frac{T-T_c(H)}{(HT)^{\frac{2}{3}}}\right].
\end{equation}
It measures the temperature shift with respect to $T_c(H)$ normalized by the fluctuation broadening~\cite{Koshelev19} and
\begin{equation}\label{Eq4b}
r_T=\left(\frac{2H_{c2}'^{2}T_c}{Gi}\right)^{1/3}
\end{equation}
is a temperature- and field-independent constant.

In Ginzburg-Landau theory, $\delta C_{mf}(T,H)$ is temperature independent. However, similar to Nb~\cite{Farrant75}, $\delta C_{mf}(T,H)$ has a sizable temperature variation in the transition region as illustrated in Fig.~\ref{Fig1}. Therefore, we have normalized our measurements to $\delta C_{mf}(T,H)$ rather than the mean-field discontinuity since we are only concerned with that part of the temperature dependence ascribed to fluctuations~\cite{Farrant75}. Here, we have determined $C_{mf}(T,H)$ for each field by fitting the data of Fig.~\ref{Fig1} to a second-order polynomials for $T<<T_m(H)$ and have extrapolated it through the transition region. The same procedure is employed to evaluate $\alpha_{mf}(T,H)$ for the cooling curves of Sample 2.   

In Fig.~\ref{Fig4}, we compare our scaled specific-heat data to the calculations of Li and Rosenstein~\cite{Li01,Li01-2,Li03,Li04,Rosenstein10} (thick solid line) who successfully derived an analytical expression of the 3d-LLL scaling function for $-25<a_T<8$, which includes the expected contributions from vortex melting. This expression was found to describe the broadening of the calorimetric transitions and the melting discontinuity in RbEuFe$_4$As$_4$~\cite{Koshelev19} extremely well. An excellent agreement with the Li-Rosenstein calculation is also achieved in both our specific-heat and thermal-expansion data for a large range of field with the constants $r_{T\perp}$ = 60 K$^{-1/3}$T$^{2/3}$ and $r_{T\parallel}$ = 160 K$^{-1/3}$T$^{2/3}$ which lead to $\tilde{\Gamma}=(r_{T\parallel}/r_{T\perp})^{\frac{3}{2}} = 4.3$. These values are in very good agreement with the respective values 68, 185 and 4.5 calculated using Eq.(\ref{Eq4b}) and the values given in Table~\ref{Table1}, demonstrating the pertinence of our scaling analysis. We note that the 3d-LLL scaling breaks down for large field values because $H_{c2}'$ in Eq.(\ref{Eq4b}) is no longer T-independent at high fields because higher-order gradient terms in the Ginzburg-Landau functional, neglected in the 3d-LLL approximation, become important.
\begin{figure}[t]
\centering
\begin{minipage}{1\columnwidth}
\centering
\includegraphics[width=1\columnwidth]{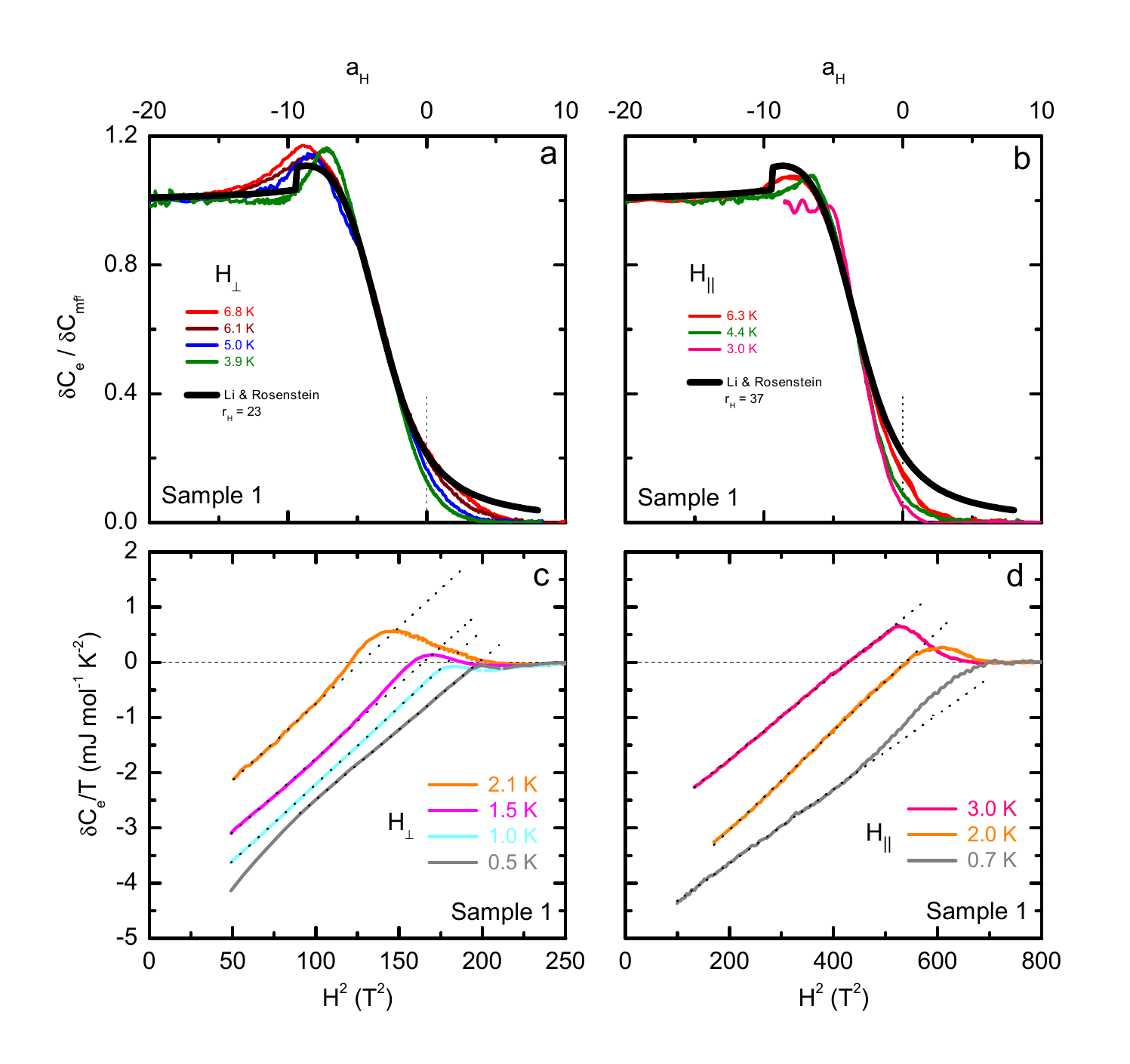}
\end{minipage}
\caption{(Color online) (a)-(b) 3d-LLL scaling of the H-dependent specific heat of Sample 1 for H $\perp$ and H $\parallel$ to the FeSe layers, respectively. The thick line represents the scaling function calculated by Li and Rosenstein~\cite{Li01,Li01-2,Li03,Li04,Rosenstein10} for the given values of $r_{H}$ (see text for details). The discontinuity at $a_H$ = -9.5 corresponds to the vortex-lattice melting transition. The dashed line indicates the mean-field $H_{c2}$(T). (c)-(d) Field dependence of $\delta C_e(H,T)/T$, the difference between the mixed- and the normal-state specific heats of Sample 1 plotted as a function $H^2$ for both field orientations.}
\label{Fig5new}
\end{figure}

A similar scaling approach should also work at very low temperatures for field curves at constant temperature. We find that a similar scaling function can be employed to analyze the mixed-state specific heat shown in Fig.~\ref{Fig3}(a). The argument of the scaling function is now
\begin{equation}\label{Eq6}
a_H=r_H\left[\frac{H-H_{c2}(T)}{(HT)^{\frac{2}{3}}}\right],
\end{equation}
with the constant
\begin{equation}\label{Eq7}
r_H=\left(\frac{2T_c}{H_{c2}'Gi}\right)^{\frac{1}{3}}.
\end{equation}
Here, we estimated C$_{mf}$(T,H) for each temperature by fitting our data to H$^2$ away from H$_m$(T) (see dotted lines in Figs~\ref{Fig5new}(c)-(d)), which is characteristic of Pauli-limited superconductors. Our scaled specific-heat data are compared to the Li-Rosenstein~\cite{Li01,Li01-2,Li03,Li04,Rosenstein10} calculation in Figs~\ref{Fig5new}(a)-(b). For $H$ $\perp$ FeSe layers, we find that our scaled data precisely collapse on the theoretical curve obtained with $r_{H \perp}$ = 23 K$^{\frac{2}{3}}$ T$^{-\frac{1}{3}}$ calculated using Eq.(\ref{Eq7}) and the values given in Table~\ref{Table1}. This agreement confirms the robustness of our scaling analysis and the existence of Gaussian thermal fluctuations in FeSe.

However, for T $<$ 6 K, the 3d-LLL scaling breaks down for $H$ $\parallel$ FeSe layers as illustrated in Fig.~\ref{Fig5new}(b). We show in Sec.~\ref{Hparallel} that it is related to strong paramagnetic effects, which are not accounted for in the 3d-LLL scaling approach.

Further, we note that the mid-point of our broad melting discontinuity lies around $a_{T}\approx-11$ {\it i.e.} below the Li-Rosenstein value~\cite{Li01,Li01-2,Li03,Li04,Rosenstein10} ($a_{T}=-9.5$) calculated for an ideal vortex lattice. We ascribe this difference to the weak flux pinning observed in our thermal-expansion measurements.

As explained in Ref.~\onlinecite{Mikitik03}, the influence of disorder on the locus of the melting line can be quantified by the parameter $D/c_L$ with $D\approx ( j_c / j_0)^{1/2}$ ($j_c $ and $j_0$ are the zero-field critical-current and depairing-current densities, respectively) and $c_L$  the Lindemann number. Using $j_c$ $\approx$ 3 $\times$ 10$^4$ A cm$^{-2}$ inferred from Ref.~\onlinecite{Sun15}, $j_0 \approx 10^{7}$ A cm$^{-2}$ and $c_L=0.2$ (see Sec.~\ref{Hperp}), we obtain $D/c_L \approx 0.3$ $<<$ 1 indicating that the observed melting line in FeSe lies very close to the genuine first-order transition line of the defect-free sample~\cite{Mikitik03}.           

\begin{figure}[b]
\centering
\begin{minipage}{1\columnwidth}
\centering
\includegraphics[width=1\columnwidth]{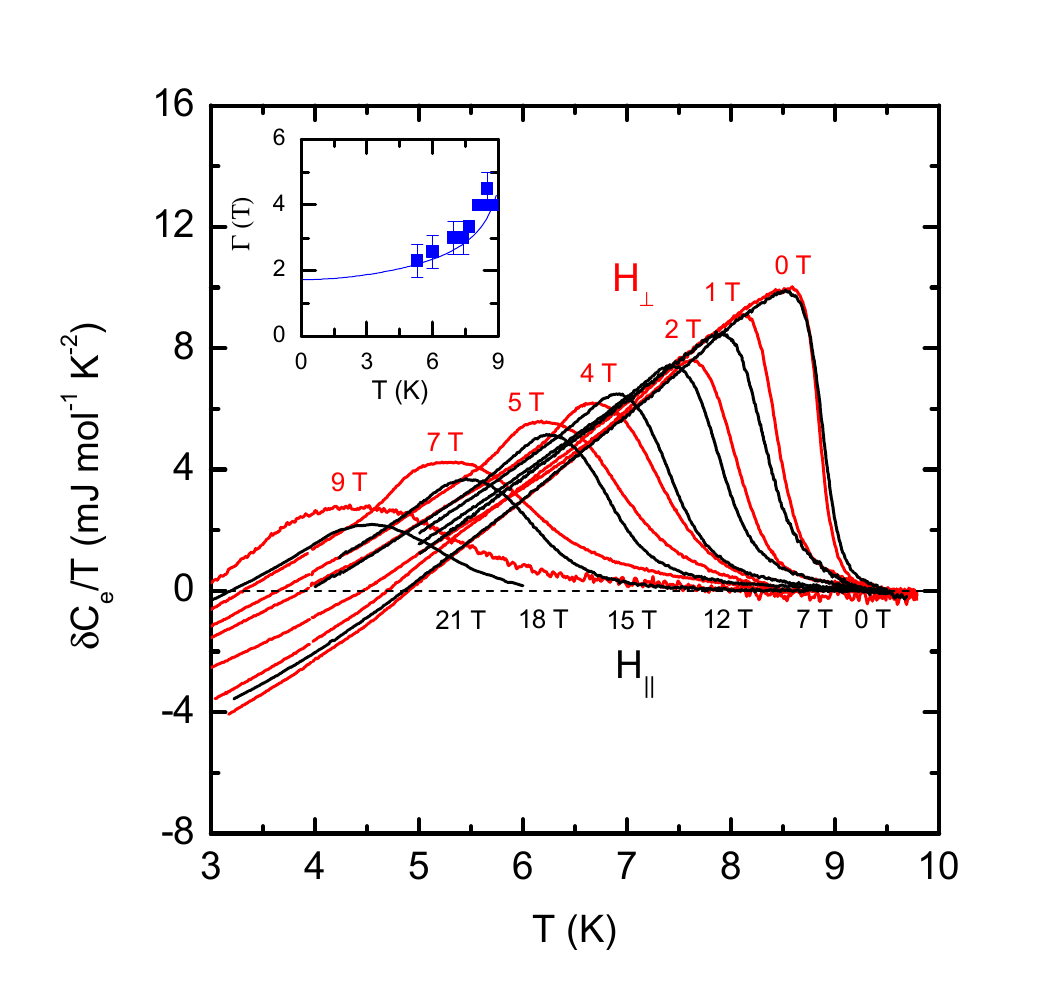}
\end{minipage}
\caption{(Color online) Comparison of the specific-heat curves measured for H perpendicular (red) and parallel (black) to the FeSe layers. The inset shows an estimate of the temperature dependence of the $H_{c2}(T)$ anisotropy. The line represents the ratio $H_{c2 \parallel}(T)/H_{c2 \perp}(T)$ obtained in Sec.~\ref{phase}}.
\label{Fig5}
\end{figure}

Hereabove, we have employed expressions derived for a single-band system whereas FeSe is a two-band superconductor. We believe that it is a fair approximation since the $k$-averaged energy gaps on the electron and hole bands are found almost equal, {\it i.e.} $\left< \Delta_h (\bf k)\right>_k \approx \left< \Delta_e (\bf k)\right>_k$ $\approx$ 1.3 meV.~\cite{Sprau17,Hardy19} In the opposite case, {\it e.g.} $\Delta_h>>\Delta_e$, 3d-LLL scaling breaks down because of the existence of two distinct energy modes~\cite{Koshelev14,Adachi16}.


\begin{figure}[b]
\centering
\begin{minipage}{0.85\columnwidth}
\centering
\includegraphics[width=1\columnwidth]{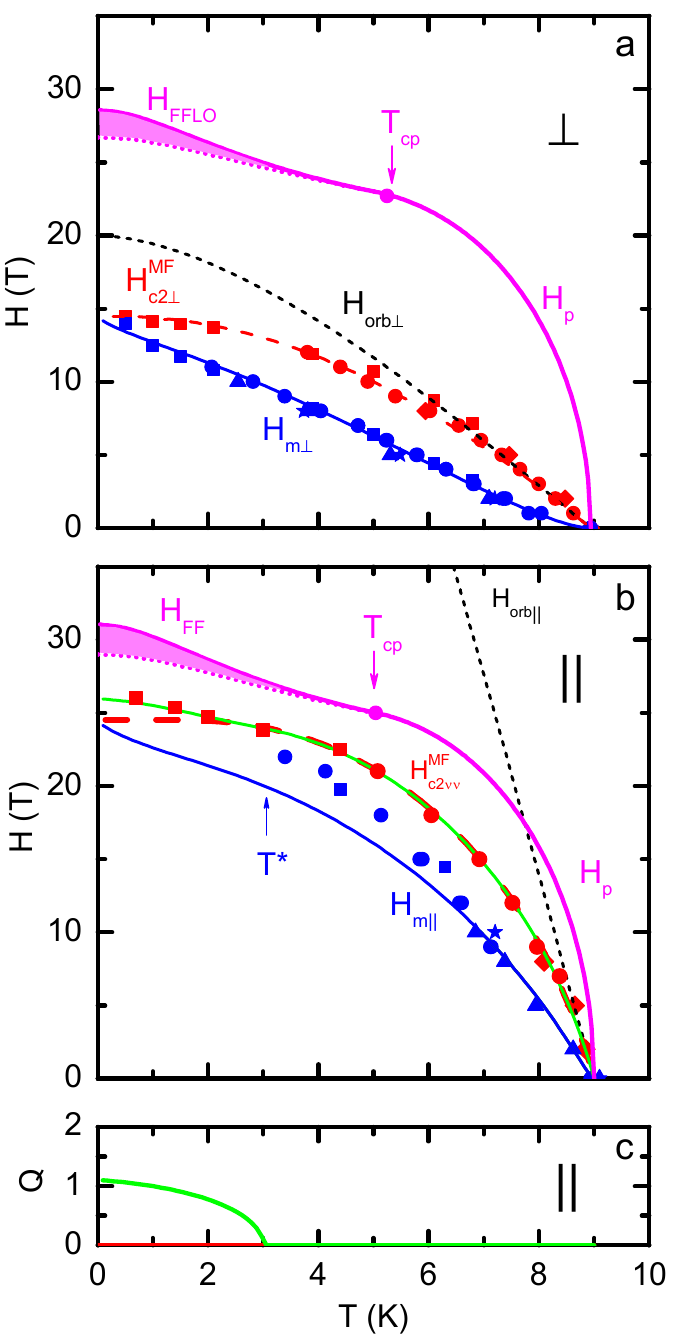}
\end{minipage}
\caption{(Color online) (a)-(b) $H-T$ phase diagram of FeSe for H $\perp$ and $\parallel$ FeSe layers, respectively. Blue (red) symbols represent the melting (mean-field upper critical field H$_{c2}$(T)) data. Black dotted lines are calculations of the orbital critical field H$_{orb}$(T). Dashed red lines represent fits to the H$_{c2}$(T) data including Pauli limitation (assuming a second-order transition). The solid blue lines represent calculation of the melting within the Lindemann approximation (see text). T$^*$ marks the temperature where H$_m$(T) and H$_{c2}$(T) merge. The green solid line is the same as the dashed red line allowing for a FFLO modulation $Q$(T) (shown in (c)), assuming a second-order phase transition between the FFLO and the normal states. For comparison, the solid and dotted magenta line are second- and first-order transitions calculated for the pure Pauli-limited case. The shaded area denotes the extent of the FFLO region in the $H-T$ plane. T$_{cp}$ stands for the tricritical point.  }
\label{Fig6}
\end{figure}

\subsection{$H-T$ phase diagram}\label{phase}
 
The above scaling approach yields a mean-field H$_{c2}$(T) and, together with the position of the melting anomaly, allows us to construct ($H-T$) phase diagrams which are displayed in Figs~\ref{Fig6}(a)-(b) for H $\perp$ and $\parallel$ to the FeSe layers, respectively. The blue symbols correspond to the vortex-melting line obtained from $T-$ (circle) and $H-$sweep (square) specific-heat measurements and triangles are inferred from thermal expansion. The locus of the mean-field $H_{c2}(T)$, derived from our scaling analysis, is represented by the red symbols and corresponds to a$_T$, a$_H$ = 0 as indicated by the dotted lines in Figs~\ref{Fig4} and~\ref{Fig5new}. In the following, we critically analyze these thermodynamically-derived phase diagrams, which are more representative of the real phase transitions than transport-derived phase diagrams~\cite{Korean20}, since the zero-resistance criteria of the latter only mark the vortex melting line.

\subsubsection{H $\perp$ FeSe layers}\label{Hperp}
 
The large value of the Ginzburg-Landau parameter $\kappa_{\perp}\approx 100$ clearly indicates that FeSe is a strong type II superconductor and thus represents a good candidate to study the influence of Pauli-depairing effects on the vortex state. The importance of spin paramagnetism is typically quantified by the Maki parameter~\cite{Maki64}

\begin{equation}\label{Eq8}
\alpha_{M} = \sqrt{2}\frac{H_{orb}(0)}{H_{p}(0)},
\end{equation}
where $H_{p}(0)=\frac{\sqrt{2}\Delta(0)}{g\mu_B}$ and $H_{orb}(0)=-0.727H_{c2}'T_c$ are the zero-temperature Pauli and orbital critical fields,~\cite{Sarma63,Helfand66} respectively ($g$ is the gyromagnetic factor). In Fig.~\ref{Fig6}(a), we show the temperature dependence of H$_{orb}(T)$ computed within the clean-limit Helfand-Werthamer framework~\cite{Helfand66} using the measured initial slope $H_{c2\perp}'=-3$ T K$^{-1}$ (black dotted lines). We obtain $H_{orb\perp}(0)=20$ T which clearly exceeds the experimental value of H$_{c2\perp}(0)\approx$ 15 T by a significant amount, strongly suggesting that Pauli depairing is already significant in FeSe for H $\perp$ FeSe layers. To account for this effect, we analyze our data within the Werthamer-Helfand-Hohenberg (WHH) formalism~\cite{Werthamer66,Brison95} including Pauli effects. An excellent fit to our data (red dashed line) is obtained for H$_{p}(0)$ = 26.5 T (left as a free parameter) which leads to a moderate value of the Maki parameter $\alpha_{M\perp}\approx$ 1.1. For completeness, we also plot in Fig.~\ref{Fig6} the temperature dependence of H$_p$(T) (magenta curve)~\cite{Sarma63}.

The melting line $H_{m\perp}$(T) (blue symbols) exhibits a characteristic upward curvature near $T_c$, similar to that of YBCO,~\cite{Bouquet01} and then crosses over to a quasi-linear dependence at lower temperature, where it finally merges with $H_{c2}$(T) at T $\approx$ 0. The observation of the vortex-melting transition down to $T/T_{c}\approx 0.1$ in FeSe allows us to examine the numerous theoretical models put forward to describe this phenomenon. Here, we compare our results to the semiquantitative approach of Houghton {\it et al.}~\cite{Houghton89} based on the Lindemann criterion. In this approach, the flux-line lattice melts when the mean-square amplitude $u^2_{th}$ of the fluctuations  
\begin{equation}\label{Lindemann}
u^2_{th}=a_0^2 \beta_A \frac{\sqrt{Gi}}{2\pi} \frac{t}{1-b_m} \sqrt{\frac{b_m}{b_{c2}}} F_T(b_m) \approx c_L^2 a{_0}^2,
\end{equation}
where $a_0\sim \Phi_0/B$ is the vortex-lattice spacing, $\beta_A$ is the Abrikosov parameter, $b_m=H_m(T)/H_{c2}(T)$, $b_{c2}=H_{c2}(T)/H_{c2}(0)$ and $t=T/T_c$. Here, $c_L\approx 0.1-0.2$ is the Lindemann number and the analytical expression of $F_T$ is given in Ref.~\onlinecite{Babich94}. We note that, for $b_m \approx 1$, Eq.(\ref{Lindemann}) leads to $r_H[H_{c2}(t)-H_m(t)]/[H_mT]^{2/3} \approx cste$ which coincides with Eq.(\ref{Eq6}) for H = H$_m$~\cite{Mikitik03}. Using our WHH calculation for H$_{c2}$(T) and $Gi$ from Table~\ref{Table1}, we solve Eq.(\ref{Lindemann}) and the resulting curve is depicted in Fig.~\ref{Fig6}(a) (solid blue line). For c$_L$ = 0.15, our calculation fit our data remarkably well down to T = 0. Thus, the melting transition in the presence of weak/moderate paramagnetic depairing remains the only genuine thermodynamic phase transition for H $\perp$ FeSe layers, as emphasized by Adachi and Ikeda~\cite{Adachi03}, and the vortex liquid smoothly crosses over to the nematic state near H$_{c2}$(T).  Finally, we note that we find no thermodynamic evidence of the high-field modulated phase reported by Kasahara {\it et al.}~\cite{Kasahara14} and Watashige {\it et al.}~\cite{Watashige17} inferred from heat- and electrical-transport and torque measurements.

\subsubsection{H $\parallel$ FeSe layers}\label{Hparallel}
The situation is very different for H $\parallel$ FeSe layers since paramagnetic effects are much stronger in this direction due to the larger value of $\alpha_{M\parallel}$=$\tilde{\Gamma}\alpha_{M\perp}\approx$ 4.5. In such a case, Adachi and Ikeda~\cite{Adachi03} predicted that H$_m$(T) and H$_{c2}$(T) could already merge at finite temperature. This appears to be the case realized in our data, as illustrated in Fig.~\ref{Fig6}(b) where the two lines merge around T$^{*}$ $\approx$ 3 K. Below this temperature, the normal state is recovered before the vortex solid had a chance to melt due to strong paramagnetic pair-breaking, which strongly suppresses H$_{c2}$(T). The vortex liquid phase thus no longer exists in this part of the phase diagram, and H$_{c2}$(T) again turns into a genuine phase transition for T $<$ T$^{*}$. This reveals that paramagnetic effects tend to suppress superconducting fluctuations, as can be expected as the transition becomes first-order like. We note that the progressive disappearance of the vortex-melting anomaly nicely correlates with the drastic reduction of the resistive transition width reported by Kasahara {\it et al}~\cite{Kasahara19}.   

In Fig.~\ref{Fig6}(b), we show our calculation of H$_{c2}$(T) (red dashed curve) using the WHH formula with the inferred values of $\alpha_{M\parallel}$ = 4.5 and $H_{orb\parallel}'=\tilde{\Gamma} H_{c2\perp}'$ = - 13.8 T K$^{-1}$. We find that it accurately reproduces our H$_{c2}$(T) data only for T $>$ T$^{*}$. For T $<$ T$^{*}$, the H$_{c2}$(T) values lie above this line, and the behavior is reminiscent of FFLO-type physics. The vortex melting line (solid blue curve), obtained within the Lindemann approximation with the same value of $Gi$ (see Sec.~\ref{Hperp}) and a slightly higher c$_{L}$ = 0.2, describes our experimental data quite well only for T $>$ 6.5 K. Below 6.5 K, the real melting (blue symbols) deviates strongly from this line, which is not unexpected, since strong paramagnetic effects are not accounted for in this model.  We note that the 'jump' observed by Kasahara {\it et al}~\cite{Kasahara19} in heat-transport measurements appears to coincide with the dashed red line at low T. Theoretically, this line no longer represents a genuine phase transition for T $<$ T$^{*}$, as explained in Refs ~\onlinecite{Sarma63,Sarma}, and the correlation with our experiments is unclear. 

In purely Pauli-limited superconductors, a spatially-modulated phase is predicted to appear at high field for T $<$ T$_{cp}$ = 0.56 T$_{c}$, as shown independently by Fulde and Ferrell $\left(\textrm{FF}, \Delta ({\bf q})e^{i{\bf qr}}\right)$ and Larkin and Ovchinnikov $\left(\textrm{LO}, \Delta\left({\bf r}\right)\cos \left({\bf qr}\right)\right)$~\cite{Fulde64,Larkin65}. Two effects are already expected to mark the emergence of the FFLO phase below T$_{cp}$: (i) a first-order transition between the uniform BCS and the modulated states (dotted magenta line) and (ii) an enhancement of H$_{c2}$(T) which now defines the transition between the FFLO- and the polarized normal-state phases (solid magenta line). However, these lines are expected to lie very close to each other in purely Pauli-limited 3D systems. Here, they are only 2 T apart at T = 0 (see shaded area in Fig.~\ref{Fig6}(b)). Accounting for the orbital effect, we expect a shift of T$_{cp}$ towards lower temperature and the two lines to lie even closer to each other~\cite{Gruenberg66,Houzet01,Houzet06,Houzet07}. For $\alpha_{M\parallel}$ $\approx$ 4.5, the FFLO state should emerge theoretically below T$_{cp}$ = 0.33 T$_c$ $\approx$ 3 K in FeSe~\cite{Brison97}. Interestingly, this corresponds to T$^{*}$ where H$_{m}$(T) and H$_{c2}$(T) are found to merge. Therefore, we have recalculated H$_{c2}$(T) (without additional parameters) allowing for a finite modulation of the order parameter, $Q = \hbar v_{F}q/2T_c$, assuming a second-order FFLO/N transition~\cite{Buzdin96}. The results for H$_{c2}(T)$ and the related Q(T) are displayed in Figs~\ref{Fig6}(b)-(c) (green solid lines), respectively. They reproduce the enhancement of H$_{c2}$(T) observed for T < T$^{*}$ remarkably well. Unfortunately, we found no experimental evidence for the required BCS/FFLO transition in our specific-heat measurements. However, recent heat-transport data from Kasahara {\it et al.}~\cite{Kasahara19} provide some evidence for such a modulated phase for T < 2 K.

It appears that the transition to the normal state possibly changes its character below T*. While it appears second -order like at T = 3 K, the transition exhibits a rather broad specific-heat discontinuity at T = 0.7 K (see Fig.~\ref{Fig3}(d)). This feature is reminiscent of the broadened discontinuity observed around $\approx$ 27 T in heat-transport measurements~\cite{Kasahara19}. Overall our phase diagram is found to be in rough agreement with that of Kasahara {\it et al.}~\cite{Kasahara19}. However, additional measurements ({\it e.g.} magnetocaloric measurements) would be very useful to further establish the firm existence of an FFLO phase in FeSe.

\subsubsection{Consistency check of our analysis}\label{Consistency}
It is worth noting that the analysis presented in Sec.~\ref{Hperp} and~\ref{Hparallel}, for both field orientations, were carried out with only two free parameters {\it i.e.} H$_{p}$(0) = 26-29 T and $c_L$ = 0.15-0.20 for determining H$_{c2}$(T) and H$_m$(T), respectively. The other quantities reported in Table~\ref{Table1} are directly inferred from our measurements using standard thermodynamic relations. Similarly, the agreement between the experimental values of $r_T$ and $r_H$ in Sec.\ref{Scaling} with these calculated directly from Eqs (\ref{Eq4b}) and (\ref{Eq7}) is better than 15 \%.

Further, the transition at T = 0 in the purely paramagnetic case occurs when the polarization energy equals the condensation energy, {\it i.e.} for~\cite{Sarma} : $(\chi_n-\chi_s)H^2_p(0) = H^2_{c}(0)$ where $\chi_n=(\mu_0/2)(g\mu_B)^2 N(0)$ represents the normal-state Pauli susceptibility and $\chi_s$ = 0 for a singlet superconductor. Using our values of H$_p$(0), H$_{c}$(0) = 0.12 T from our $\delta C_e$ (T,H=0) data and $\Delta(0)$ = 1.3 meV from BQPI experiments, we obtain $\chi_{n}$ $\approx$ 1.7$\times$10$^{-5}$ and $N(0)$ $\approx$ 2.2$\times$10$^{47}$ J$^{-1}$ m$^{-3}$ spin$^{-1}$. These values lead to a value of the Sommerfeld coefficient $\gamma_n=\frac{2 \pi^2}{3}k^2_{B}N(0) \approx 6.7$ mJ mol$^{-1}$ K$^{-2}$ in excellent agreement with the value inferred from direct specific-heat measurements.  These consistency checks provide us with great confidence concerning the relevance of our scaling analysis, the accuracy of the inferred mean-field H$_{c2}$(T), and the validity of the presented phase diagrams.

\section{Conclusions}\label{Conclusion}
We have determined the full $H-T$ phase diagram of the nematic superconductor FeSe for both field orientations. Compelling evidence of an underlying vortex-melting transition is found in both specific-heat and thermal-expansion measurements down to low temperature and high magnetic fields. We demonstrate the existence of significant Gaussian thermal fluctuations via a scaling analysis of our thermodynamic data which yields the temperature dependence of the mean-field upper critical field. The antagonist interplay between superconducting fluctuations and Pauli depairing effects is studied. We argue that the predominance of the paramagnetic limitation at low temperature is responsible for the unusual disappearance of the melting transition at finite temperature, around T$^{*}$ $\approx$ 2 K, for H $\parallel$ FeSe planes, as anticipated theoretically. A slight upturn of H$_{c2}$(T) for T $<$ T$^{*}$, possibly related to the occurence of the Fulde-Ferrell-Larkin-Ovchinnikov phase, is observed. Additional thermodynamic measurements {\it e.g.} of the magnetocaloric effect or magnetostriction, with accurate in-plane field alignment, are necessary to firmly establish the existence of this modulated phase.

\begin{acknowledgments}
We thank B. Rosenstein for providing us with the theoretical scaling function. We acknowledge fruitful discussions with G. P. Mikitik, M. Houzet, A. E. Koshelev, J.-P. Brison and R. Eder. Part of this work was performed at the LNCMI, a member of the European Magnetic Field Laboratory (EMFL). K.W. acknowledges funding from the Alexander von Humboldt Foundation. The contribution from M. M. was supported by the Karlsruhe Nano Micro Facility (KNMF). 
\end{acknowledgments}

\bibliography{Biblio.bib}

\begin{thebibliography}{115}%
\makeatletter
\providecommand \@ifxundefined [1]{%
 \@ifx{#1\undefined}
}%
\providecommand \@ifnum [1]{%
 \ifnum #1\expandafter \@firstoftwo
 \else \expandafter \@secondoftwo
 \fi
}%
\providecommand \@ifx [1]{%
 \ifx #1\expandafter \@firstoftwo
 \else \expandafter \@secondoftwo
 \fi
}%
\providecommand \natexlab [1]{#1}%
\providecommand \enquote  [1]{``#1''}%
\providecommand \bibnamefont  [1]{#1}%
\providecommand \bibfnamefont [1]{#1}%
\providecommand \citenamefont [1]{#1}%
\providecommand \href@noop [0]{\@secondoftwo}%
\providecommand \href [0]{\begingroup \@sanitize@url \@href}%
\providecommand \@href[1]{\@@startlink{#1}\@@href}%
\providecommand \@@href[1]{\endgroup#1\@@endlink}%
\providecommand \@sanitize@url [0]{\catcode `\\12\catcode `\$12\catcode
  `\&12\catcode `\#12\catcode `\^12\catcode `\_12\catcode `\%12\relax}%
\providecommand \@@startlink[1]{}%
\providecommand \@@endlink[0]{}%
\providecommand \url  [0]{\begingroup\@sanitize@url \@url }%
\providecommand \@url [1]{\endgroup\@href {#1}{\urlprefix }}%
\providecommand \urlprefix  [0]{URL }%
\providecommand \Eprint [0]{\href }%
\providecommand \doibase [0]{http://dx.doi.org/}%
\providecommand \selectlanguage [0]{\@gobble}%
\providecommand \bibinfo  [0]{\@secondoftwo}%
\providecommand \bibfield  [0]{\@secondoftwo}%
\providecommand \translation [1]{[#1]}%
\providecommand \BibitemOpen [0]{}%
\providecommand \bibitemStop [0]{}%
\providecommand \bibitemNoStop [0]{.\EOS\space}%
\providecommand \EOS [0]{\spacefactor3000\relax}%
\providecommand \BibitemShut  [1]{\csname bibitem#1\endcsname}%
\let\auto@bib@innerbib\@empty
\bibitem [{\citenamefont {Abrikosov}(1957)}]{Abrikosov57}%
  \BibitemOpen
  \bibfield  {author} {\bibinfo {author} {\bibfnamefont {A.~A.}\ \bibnamefont
  {Abrikosov}},\ }\href@noop {} {\bibfield  {journal} {\bibinfo  {journal}
  {Soviet Phys. JETP}\ }\textbf {\bibinfo {volume} {5}},\ \bibinfo {pages}
  {1174} (\bibinfo {year} {1957})}\BibitemShut {NoStop}%
\bibitem [{\citenamefont {de~Gennes}(1999)}]{deGennes}%
  \BibitemOpen
  \bibfield  {author} {\bibinfo {author} {\bibfnamefont {P.-G.}\ \bibnamefont
  {de~Gennes}},\ }\href@noop {} {\emph {\bibinfo {title} {Superconductivity of
  metals and alloys}}},\ Advanced book classics\ (\bibinfo  {publisher}
  {Westview, Perseus Books},\ \bibinfo {year} {1999})\BibitemShut {NoStop}%
\bibitem [{\citenamefont {Larkin}\ and\ \citenamefont
  {Ovchinnikov}(1970)}]{Larkin70}%
  \BibitemOpen
  \bibfield  {author} {\bibinfo {author} {\bibfnamefont {A.~I.}\ \bibnamefont
  {Larkin}}\ and\ \bibinfo {author} {\bibfnamefont {Y.~N.}\ \bibnamefont
  {Ovchinnikov}},\ }\href@noop {} {\bibfield  {journal} {\bibinfo  {journal}
  {Soviet Phys. JETP}\ }\textbf {\bibinfo {volume} {31}},\ \bibinfo {pages}
  {784} (\bibinfo {year} {1970})}\BibitemShut {NoStop}%
\bibitem [{\citenamefont {Larkin}\ and\ \citenamefont
  {Ovchinnikov}(1974)}]{Larkin74}%
  \BibitemOpen
  \bibfield  {author} {\bibinfo {author} {\bibfnamefont {A.~I.}\ \bibnamefont
  {Larkin}}\ and\ \bibinfo {author} {\bibfnamefont {Y.~N.}\ \bibnamefont
  {Ovchinnikov}},\ }\href@noop {} {\bibfield  {journal} {\bibinfo  {journal}
  {Soviet Phys. JETP}\ }\textbf {\bibinfo {volume} {38}},\ \bibinfo {pages}
  {854} (\bibinfo {year} {1974})}\BibitemShut {NoStop}%
\bibitem [{\citenamefont {Larkin}\ and\ \citenamefont
  {Ovchinnikov}(1979)}]{Larkin79}%
  \BibitemOpen
  \bibfield  {author} {\bibinfo {author} {\bibfnamefont {A.~I.}\ \bibnamefont
  {Larkin}}\ and\ \bibinfo {author} {\bibfnamefont {Y.~N.}\ \bibnamefont
  {Ovchinnikov}},\ }\href@noop {} {\bibfield  {journal} {\bibinfo  {journal}
  {J. Low Temp. Phys.}\ }\textbf {\bibinfo {volume} {34}},\ \bibinfo {pages}
  {409} (\bibinfo {year} {1979})}\BibitemShut {NoStop}%
\bibitem [{\citenamefont {Giamarchi}\ and\ \citenamefont
  {Le~Doussal}(1994)}]{Giamarchi94}%
  \BibitemOpen
  \bibfield  {author} {\bibinfo {author} {\bibfnamefont {T.}~\bibnamefont
  {Giamarchi}}\ and\ \bibinfo {author} {\bibfnamefont {P.}~\bibnamefont
  {Le~Doussal}},\ }\href@noop {} {\bibfield  {journal} {\bibinfo  {journal}
  {Phys. Rev. Lett.}\ }\textbf {\bibinfo {volume} {72}},\ \bibinfo {pages}
  {1530} (\bibinfo {year} {1994})}\BibitemShut {NoStop}%
\bibitem [{\citenamefont {Giamarchi}\ and\ \citenamefont
  {Le~Doussal}(1997)}]{Giamarchi97}%
  \BibitemOpen
  \bibfield  {author} {\bibinfo {author} {\bibfnamefont {T.}~\bibnamefont
  {Giamarchi}}\ and\ \bibinfo {author} {\bibfnamefont {P.}~\bibnamefont
  {Le~Doussal}},\ }\href@noop {} {\bibfield  {journal} {\bibinfo  {journal}
  {Phys. Rev. B}\ }\textbf {\bibinfo {volume} {55}},\ \bibinfo {pages} {6577}
  (\bibinfo {year} {1997})}\BibitemShut {NoStop}%
\bibitem [{\citenamefont {Saint-James}\ \emph {et~al.}(1969)\citenamefont
  {Saint-James}, \citenamefont {Sarma},\ and\ \citenamefont {Thomas}}]{Sarma}%
  \BibitemOpen
  \bibinfo {editor} {\bibfnamefont {D.}~\bibnamefont {Saint-James}}, \bibinfo
  {editor} {\bibfnamefont {G.}~\bibnamefont {Sarma}}, \ and\ \bibinfo {editor}
  {\bibfnamefont {E.~J.}\ \bibnamefont {Thomas}},\ eds.,\ \href@noop {} {\emph
  {\bibinfo {title} {Type II superconductivity}}},\ International series of
  monographs in natural philosophy ; 17\ (\bibinfo  {publisher} {Pergamon
  Press},\ \bibinfo {year} {1969})\BibitemShut {NoStop}%
\bibitem [{\citenamefont {Blatter}\ \emph
  {et~al.}(1994{\natexlab{a}})\citenamefont {Blatter}, \citenamefont
  {Feigel'man}, \citenamefont {Geshkenbein}, \citenamefont {Larkin},\ and\
  \citenamefont {Vinokur}}]{Blatter94}%
  \BibitemOpen
  \bibfield  {author} {\bibinfo {author} {\bibfnamefont {G.}~\bibnamefont
  {Blatter}}, \bibinfo {author} {\bibfnamefont {M.~V.}\ \bibnamefont
  {Feigel'man}}, \bibinfo {author} {\bibfnamefont {V.~B.}\ \bibnamefont
  {Geshkenbein}}, \bibinfo {author} {\bibfnamefont {A.~I.}\ \bibnamefont
  {Larkin}}, \ and\ \bibinfo {author} {\bibfnamefont {V.~M.}\ \bibnamefont
  {Vinokur}},\ }\href@noop {} {\bibfield  {journal} {\bibinfo  {journal} {Rev.
  Mod. Phys.}\ }\textbf {\bibinfo {volume} {66}},\ \bibinfo {pages} {1125}
  (\bibinfo {year} {1994}{\natexlab{a}})}\BibitemShut {NoStop}%
\bibitem [{\citenamefont {Brandt}(1995)}]{Brandt95}%
  \BibitemOpen
  \bibfield  {author} {\bibinfo {author} {\bibfnamefont {E.~H.}\ \bibnamefont
  {Brandt}},\ }\href@noop {} {\bibfield  {journal} {\bibinfo  {journal} {Rep.
  Prog. Phys.}\ }\textbf {\bibinfo {volume} {58}},\ \bibinfo {pages} {1465}
  (\bibinfo {year} {1995})}\BibitemShut {NoStop}%
\bibitem [{\citenamefont {Fisher}\ \emph {et~al.}(2007)\citenamefont {Fisher},
  \citenamefont {Gordon},\ and\ \citenamefont {Phillips}}]{SchriefferBook}%
  \BibitemOpen
  \bibfield  {author} {\bibinfo {author} {\bibfnamefont {R.~A.}\ \bibnamefont
  {Fisher}}, \bibinfo {author} {\bibfnamefont {J.~E.}\ \bibnamefont {Gordon}},
  \ and\ \bibinfo {author} {\bibfnamefont {N.~E.}\ \bibnamefont {Phillips}},\
  }in\ \href@noop {} {\emph {\bibinfo {booktitle} {Handbook of High-Temperature
  Superconductivity}}},\ \bibinfo {editor} {edited by\ \bibinfo {editor}
  {\bibfnamefont {J.~R.}\ \bibnamefont {Schrieffer}}\ and\ \bibinfo {editor}
  {\bibfnamefont {J.~S.}\ \bibnamefont {Brooks}}}\ (\bibinfo  {publisher}
  {Springer},\ \bibinfo {year} {2007})\ pp.\ \bibinfo {pages}
  {345--397}\BibitemShut {NoStop}%
\bibitem [{\citenamefont {Eilenberger}(1967)}]{Eilenberger67}%
  \BibitemOpen
  \bibfield  {author} {\bibinfo {author} {\bibfnamefont {G.}~\bibnamefont
  {Eilenberger}},\ }\href@noop {} {\bibfield  {journal} {\bibinfo  {journal}
  {Phys. Rev.}\ }\textbf {\bibinfo {volume} {164}},\ \bibinfo {pages} {628}
  (\bibinfo {year} {1967})}\BibitemShut {NoStop}%
\bibitem [{\citenamefont {Huberman}\ and\ \citenamefont
  {Doniach}(1979)}]{Huberman79}%
  \BibitemOpen
  \bibfield  {author} {\bibinfo {author} {\bibfnamefont {B.~A.}\ \bibnamefont
  {Huberman}}\ and\ \bibinfo {author} {\bibfnamefont {S.}~\bibnamefont
  {Doniach}},\ }\href@noop {} {\bibfield  {journal} {\bibinfo  {journal} {Phys.
  Rev. Lett.}\ }\textbf {\bibinfo {volume} {43}},\ \bibinfo {pages} {950}
  (\bibinfo {year} {1979})}\BibitemShut {NoStop}%
\bibitem [{\citenamefont {Fisher}(1980)}]{Fisher80}%
  \BibitemOpen
  \bibfield  {author} {\bibinfo {author} {\bibfnamefont {D.~S.}\ \bibnamefont
  {Fisher}},\ }\href@noop {} {\bibfield  {journal} {\bibinfo  {journal} {Phys.
  Rev. B}\ }\textbf {\bibinfo {volume} {22}},\ \bibinfo {pages} {1190}
  (\bibinfo {year} {1980})}\BibitemShut {NoStop}%
\bibitem [{\citenamefont {Zeldov}\ \emph {et~al.}(1995)\citenamefont {Zeldov},
  \citenamefont {Majer}, \citenamefont {Konczykowski}, \citenamefont
  {Geshkenbein}, \citenamefont {Vinokur},\ and\ \citenamefont
  {Shtrikman}}]{Zeldov95}%
  \BibitemOpen
  \bibfield  {author} {\bibinfo {author} {\bibfnamefont {E.}~\bibnamefont
  {Zeldov}}, \bibinfo {author} {\bibfnamefont {D.}~\bibnamefont {Majer}},
  \bibinfo {author} {\bibfnamefont {M.}~\bibnamefont {Konczykowski}}, \bibinfo
  {author} {\bibfnamefont {V.~B.}\ \bibnamefont {Geshkenbein}}, \bibinfo
  {author} {\bibfnamefont {V.~M.}\ \bibnamefont {Vinokur}}, \ and\ \bibinfo
  {author} {\bibfnamefont {H.}~\bibnamefont {Shtrikman}},\ }\href@noop {}
  {\bibfield  {journal} {\bibinfo  {journal} {Nature}\ }\textbf {\bibinfo
  {volume} {375}},\ \bibinfo {pages} {373} (\bibinfo {year}
  {1995})}\BibitemShut {NoStop}%
\bibitem [{\citenamefont {Welp}\ \emph {et~al.}(1996)\citenamefont {Welp},
  \citenamefont {Fendrich}, \citenamefont {Kwok}, \citenamefont {Crabtree},\
  and\ \citenamefont {Veal}}]{Welp96}%
  \BibitemOpen
  \bibfield  {author} {\bibinfo {author} {\bibfnamefont {U.}~\bibnamefont
  {Welp}}, \bibinfo {author} {\bibfnamefont {J.~A.}\ \bibnamefont {Fendrich}},
  \bibinfo {author} {\bibfnamefont {W.~K.}\ \bibnamefont {Kwok}}, \bibinfo
  {author} {\bibfnamefont {G.~W.}\ \bibnamefont {Crabtree}}, \ and\ \bibinfo
  {author} {\bibfnamefont {B.~W.}\ \bibnamefont {Veal}},\ }\href@noop {}
  {\bibfield  {journal} {\bibinfo  {journal} {Phys. Rev. Lett.}\ }\textbf
  {\bibinfo {volume} {76}},\ \bibinfo {pages} {4809} (\bibinfo {year}
  {1996})}\BibitemShut {NoStop}%
\bibitem [{\citenamefont {Liang}\ \emph {et~al.}(1996)\citenamefont {Liang},
  \citenamefont {Bonn},\ and\ \citenamefont {Hardy}}]{Liang96}%
  \BibitemOpen
  \bibfield  {author} {\bibinfo {author} {\bibfnamefont {R.}~\bibnamefont
  {Liang}}, \bibinfo {author} {\bibfnamefont {D.~A.}\ \bibnamefont {Bonn}}, \
  and\ \bibinfo {author} {\bibfnamefont {W.~N.}\ \bibnamefont {Hardy}},\
  }\href@noop {} {\bibfield  {journal} {\bibinfo  {journal} {Phys. Rev. Lett.}\
  }\textbf {\bibinfo {volume} {76}},\ \bibinfo {pages} {835} (\bibinfo {year}
  {1996})}\BibitemShut {NoStop}%
\bibitem [{\citenamefont {Nishizaki}\ \emph {et~al.}(1996)\citenamefont
  {Nishizaki}, \citenamefont {Onodera}, \citenamefont {Kobayashi},
  \citenamefont {Asaoka},\ and\ \citenamefont {Takei}}]{Nishizaki96}%
  \BibitemOpen
  \bibfield  {author} {\bibinfo {author} {\bibfnamefont {T.}~\bibnamefont
  {Nishizaki}}, \bibinfo {author} {\bibfnamefont {Y.}~\bibnamefont {Onodera}},
  \bibinfo {author} {\bibfnamefont {N.}~\bibnamefont {Kobayashi}}, \bibinfo
  {author} {\bibfnamefont {H.}~\bibnamefont {Asaoka}}, \ and\ \bibinfo {author}
  {\bibfnamefont {H.}~\bibnamefont {Takei}},\ }\href@noop {} {\bibfield
  {journal} {\bibinfo  {journal} {Phys. Rev. B}\ }\textbf {\bibinfo {volume}
  {53}},\ \bibinfo {pages} {82} (\bibinfo {year} {1996})}\BibitemShut {NoStop}%
\bibitem [{\citenamefont {Schilling}\ \emph {et~al.}(1996)\citenamefont
  {Schilling}, \citenamefont {Fisher}, \citenamefont {Phillips}, \citenamefont
  {Welp}, \citenamefont {Dasgupta}, \citenamefont {Kwok},\ and\ \citenamefont
  {Crabtree}}]{Schilling96}%
  \BibitemOpen
  \bibfield  {author} {\bibinfo {author} {\bibfnamefont {A.}~\bibnamefont
  {Schilling}}, \bibinfo {author} {\bibfnamefont {R.~A.}\ \bibnamefont
  {Fisher}}, \bibinfo {author} {\bibfnamefont {N.~E.}\ \bibnamefont
  {Phillips}}, \bibinfo {author} {\bibfnamefont {U.}~\bibnamefont {Welp}},
  \bibinfo {author} {\bibfnamefont {D.}~\bibnamefont {Dasgupta}}, \bibinfo
  {author} {\bibfnamefont {W.~K.}\ \bibnamefont {Kwok}}, \ and\ \bibinfo
  {author} {\bibfnamefont {G.~W.}\ \bibnamefont {Crabtree}},\ }\href@noop {}
  {\bibfield  {journal} {\bibinfo  {journal} {Phys. Rev. B}\ }\textbf {\bibinfo
  {volume} {382}},\ \bibinfo {pages} {791} (\bibinfo {year}
  {1996})}\BibitemShut {NoStop}%
\bibitem [{\citenamefont {Schilling}\ \emph {et~al.}(1997)\citenamefont
  {Schilling}, \citenamefont {Fisher}, \citenamefont {Phillips}, \citenamefont
  {Welp}, \citenamefont {Kwok},\ and\ \citenamefont {Crabtree}}]{Schilling97}%
  \BibitemOpen
  \bibfield  {author} {\bibinfo {author} {\bibfnamefont {A.}~\bibnamefont
  {Schilling}}, \bibinfo {author} {\bibfnamefont {R.~A.}\ \bibnamefont
  {Fisher}}, \bibinfo {author} {\bibfnamefont {N.~E.}\ \bibnamefont
  {Phillips}}, \bibinfo {author} {\bibfnamefont {U.}~\bibnamefont {Welp}},
  \bibinfo {author} {\bibfnamefont {W.~K.}\ \bibnamefont {Kwok}}, \ and\
  \bibinfo {author} {\bibfnamefont {G.~W.}\ \bibnamefont {Crabtree}},\
  }\href@noop {} {\bibfield  {journal} {\bibinfo  {journal} {Phys. Rev. Lett.}\
  }\textbf {\bibinfo {volume} {78}},\ \bibinfo {pages} {4833} (\bibinfo {year}
  {1997})}\BibitemShut {NoStop}%
\bibitem [{\citenamefont {Schilling}\ \emph {et~al.}(1998)\citenamefont
  {Schilling}, \citenamefont {Fisher}, \citenamefont {Phillips}, \citenamefont
  {Welp}, \citenamefont {Kwok},\ and\ \citenamefont {Crabtree}}]{Schilling98}%
  \BibitemOpen
  \bibfield  {author} {\bibinfo {author} {\bibfnamefont {A.}~\bibnamefont
  {Schilling}}, \bibinfo {author} {\bibfnamefont {R.~A.}\ \bibnamefont
  {Fisher}}, \bibinfo {author} {\bibfnamefont {N.~E.}\ \bibnamefont
  {Phillips}}, \bibinfo {author} {\bibfnamefont {U.}~\bibnamefont {Welp}},
  \bibinfo {author} {\bibfnamefont {W.~K.}\ \bibnamefont {Kwok}}, \ and\
  \bibinfo {author} {\bibfnamefont {G.~W.}\ \bibnamefont {Crabtree}},\
  }\href@noop {} {\bibfield  {journal} {\bibinfo  {journal} {Phys. Rev. B}\
  }\textbf {\bibinfo {volume} {58}},\ \bibinfo {pages} {11157} (\bibinfo {year}
  {1998})}\BibitemShut {NoStop}%
\bibitem [{\citenamefont {Roulin}\ \emph
  {et~al.}(1995{\natexlab{a}})\citenamefont {Roulin}, \citenamefont {Junod},
  \citenamefont {Erb},\ and\ \citenamefont {Walker}}]{Roulin95}%
  \BibitemOpen
  \bibfield  {author} {\bibinfo {author} {\bibfnamefont {M.}~\bibnamefont
  {Roulin}}, \bibinfo {author} {\bibfnamefont {A.}~\bibnamefont {Junod}},
  \bibinfo {author} {\bibfnamefont {A.}~\bibnamefont {Erb}}, \ and\ \bibinfo
  {author} {\bibfnamefont {E.}~\bibnamefont {Walker}},\ }\href@noop {}
  {\bibfield  {journal} {\bibinfo  {journal} {J. Low Temp. Phys.}\ }\textbf
  {\bibinfo {volume} {105}},\ \bibinfo {pages} {1099} (\bibinfo {year}
  {1995}{\natexlab{a}})}\BibitemShut {NoStop}%
\bibitem [{\citenamefont {Roulin}\ \emph {et~al.}(1996)\citenamefont {Roulin},
  \citenamefont {Junod},\ and\ \citenamefont {Walker}}]{Roulin96}%
  \BibitemOpen
  \bibfield  {author} {\bibinfo {author} {\bibfnamefont {M.}~\bibnamefont
  {Roulin}}, \bibinfo {author} {\bibfnamefont {A.}~\bibnamefont {Junod}}, \
  and\ \bibinfo {author} {\bibfnamefont {E.}~\bibnamefont {Walker}},\
  }\href@noop {} {\bibfield  {journal} {\bibinfo  {journal} {Science}\ }\textbf
  {\bibinfo {volume} {273}},\ \bibinfo {pages} {1210} (\bibinfo {year}
  {1996})}\BibitemShut {NoStop}%
\bibitem [{\citenamefont {Roulin}\ \emph
  {et~al.}(1998{\natexlab{a}})\citenamefont {Roulin}, \citenamefont {Junod},\
  and\ \citenamefont {Walker}}]{Roulin98}%
  \BibitemOpen
  \bibfield  {author} {\bibinfo {author} {\bibfnamefont {M.}~\bibnamefont
  {Roulin}}, \bibinfo {author} {\bibfnamefont {A.}~\bibnamefont {Junod}}, \
  and\ \bibinfo {author} {\bibfnamefont {E.}~\bibnamefont {Walker}},\
  }\href@noop {} {\bibfield  {journal} {\bibinfo  {journal} {Physica C}\
  }\textbf {\bibinfo {volume} {296}},\ \bibinfo {pages} {137} (\bibinfo {year}
  {1998}{\natexlab{a}})}\BibitemShut {NoStop}%
\bibitem [{\citenamefont {Revaz}\ \emph {et~al.}(1998)\citenamefont {Revaz},
  \citenamefont {Junod},\ and\ \citenamefont {Erb}}]{Revaz98}%
  \BibitemOpen
  \bibfield  {author} {\bibinfo {author} {\bibfnamefont {B.}~\bibnamefont
  {Revaz}}, \bibinfo {author} {\bibfnamefont {A.}~\bibnamefont {Junod}}, \ and\
  \bibinfo {author} {\bibfnamefont {A.}~\bibnamefont {Erb}},\ }\href@noop {}
  {\bibfield  {journal} {\bibinfo  {journal} {Phys. Rev. B}\ }\textbf {\bibinfo
  {volume} {58}},\ \bibinfo {pages} {11153} (\bibinfo {year}
  {1998})}\BibitemShut {NoStop}%
\bibitem [{\citenamefont {Bouquet}\ \emph {et~al.}(2001)\citenamefont
  {Bouquet}, \citenamefont {Marcenat}, \citenamefont {Steep}, \citenamefont
  {Calemczuk}, \citenamefont {Kwok}, \citenamefont {Welp}, \citenamefont
  {Crabtree}, \citenamefont {Fisher}, \citenamefont {Phillips},\ and\
  \citenamefont {Schilling}}]{Bouquet01}%
  \BibitemOpen
  \bibfield  {author} {\bibinfo {author} {\bibfnamefont {F.}~\bibnamefont
  {Bouquet}}, \bibinfo {author} {\bibfnamefont {C.}~\bibnamefont {Marcenat}},
  \bibinfo {author} {\bibfnamefont {E.}~\bibnamefont {Steep}}, \bibinfo
  {author} {\bibfnamefont {R.}~\bibnamefont {Calemczuk}}, \bibinfo {author}
  {\bibfnamefont {W.~K.}\ \bibnamefont {Kwok}}, \bibinfo {author}
  {\bibfnamefont {U.}~\bibnamefont {Welp}}, \bibinfo {author} {\bibfnamefont
  {G.~W.}\ \bibnamefont {Crabtree}}, \bibinfo {author} {\bibfnamefont {R.~A.}\
  \bibnamefont {Fisher}}, \bibinfo {author} {\bibfnamefont {N.~E.}\
  \bibnamefont {Phillips}}, \ and\ \bibinfo {author} {\bibfnamefont
  {A.}~\bibnamefont {Schilling}},\ }\href@noop {} {\bibfield  {journal}
  {\bibinfo  {journal} {Nature}\ }\textbf {\bibinfo {volume} {411}},\ \bibinfo
  {pages} {448} (\bibinfo {year} {2001})}\BibitemShut {NoStop}%
\bibitem [{\citenamefont {Lortz}\ \emph
  {et~al.}(2003{\natexlab{a}})\citenamefont {Lortz}, \citenamefont {Meingast},
  \citenamefont {Welp}, \citenamefont {Kwok},\ and\ \citenamefont
  {Crabtree}}]{Lortz03M}%
  \BibitemOpen
  \bibfield  {author} {\bibinfo {author} {\bibfnamefont {R.}~\bibnamefont
  {Lortz}}, \bibinfo {author} {\bibfnamefont {C.}~\bibnamefont {Meingast}},
  \bibinfo {author} {\bibfnamefont {U.}~\bibnamefont {Welp}}, \bibinfo {author}
  {\bibfnamefont {W.~K.}\ \bibnamefont {Kwok}}, \ and\ \bibinfo {author}
  {\bibfnamefont {G.~W.}\ \bibnamefont {Crabtree}},\ }\href@noop {} {\bibfield
  {journal} {\bibinfo  {journal} {Phys. Rev. Lett.}\ }\textbf {\bibinfo
  {volume} {90}},\ \bibinfo {pages} {237002} (\bibinfo {year}
  {2003}{\natexlab{a}})}\BibitemShut {NoStop}%
\bibitem [{\citenamefont {Lortz}\ \emph {et~al.}(2006)\citenamefont {Lortz},
  \citenamefont {Lin}, \citenamefont {Musolino}, \citenamefont {Wang},
  \citenamefont {Junod}, \citenamefont {Rosenstein},\ and\ \citenamefont
  {Toyota}}]{Lortz06}%
  \BibitemOpen
  \bibfield  {author} {\bibinfo {author} {\bibfnamefont {R.}~\bibnamefont
  {Lortz}}, \bibinfo {author} {\bibfnamefont {F.}~\bibnamefont {Lin}}, \bibinfo
  {author} {\bibfnamefont {N.}~\bibnamefont {Musolino}}, \bibinfo {author}
  {\bibfnamefont {Y.}~\bibnamefont {Wang}}, \bibinfo {author} {\bibfnamefont
  {A.}~\bibnamefont {Junod}}, \bibinfo {author} {\bibfnamefont
  {B.}~\bibnamefont {Rosenstein}}, \ and\ \bibinfo {author} {\bibfnamefont
  {N.}~\bibnamefont {Toyota}},\ }\href@noop {} {\bibfield  {journal} {\bibinfo
  {journal} {Phys. Rev. B}\ }\textbf {\bibinfo {volume} {74}},\ \bibinfo
  {pages} {104502} (\bibinfo {year} {2006})}\BibitemShut {NoStop}%
\bibitem [{\citenamefont {Petrovi\ifmmode~\acute{c}\else \'{c}\fi{}}\ \emph
  {et~al.}(2009)\citenamefont {Petrovi\ifmmode~\acute{c}\else \'{c}\fi{}},
  \citenamefont {Fasano}, \citenamefont {Lortz}, \citenamefont {Senatore},
  \citenamefont {Demuer}, \citenamefont {Antunes}, \citenamefont {Par\'e},
  \citenamefont {Salloum}, \citenamefont {Gougeon}, \citenamefont {Potel},\
  and\ \citenamefont {Fischer}}]{Petrovic09}%
  \BibitemOpen
  \bibfield  {author} {\bibinfo {author} {\bibfnamefont {A.~P.}\ \bibnamefont
  {Petrovi\ifmmode~\acute{c}\else \'{c}\fi{}}}, \bibinfo {author}
  {\bibfnamefont {Y.}~\bibnamefont {Fasano}}, \bibinfo {author} {\bibfnamefont
  {R.}~\bibnamefont {Lortz}}, \bibinfo {author} {\bibfnamefont
  {C.}~\bibnamefont {Senatore}}, \bibinfo {author} {\bibfnamefont
  {A.}~\bibnamefont {Demuer}}, \bibinfo {author} {\bibfnamefont {A.~B.}\
  \bibnamefont {Antunes}}, \bibinfo {author} {\bibfnamefont {A.}~\bibnamefont
  {Par\'e}}, \bibinfo {author} {\bibfnamefont {D.}~\bibnamefont {Salloum}},
  \bibinfo {author} {\bibfnamefont {P.}~\bibnamefont {Gougeon}}, \bibinfo
  {author} {\bibfnamefont {M.}~\bibnamefont {Potel}}, \ and\ \bibinfo {author}
  {\bibfnamefont {O.}~\bibnamefont {Fischer}},\ }\href@noop {} {\bibfield
  {journal} {\bibinfo  {journal} {Phys. Rev. Lett.}\ }\textbf {\bibinfo
  {volume} {103}},\ \bibinfo {pages} {257001} (\bibinfo {year}
  {2009})}\BibitemShut {NoStop}%
\bibitem [{\citenamefont {Mak}\ \emph {et~al.}(2013)\citenamefont {Mak},
  \citenamefont {Burger}, \citenamefont {Cevey}, \citenamefont {Wolf},
  \citenamefont {Meingast},\ and\ \citenamefont {Lortz}}]{Mak13}%
  \BibitemOpen
  \bibfield  {author} {\bibinfo {author} {\bibfnamefont {H.~K.}\ \bibnamefont
  {Mak}}, \bibinfo {author} {\bibfnamefont {P.}~\bibnamefont {Burger}},
  \bibinfo {author} {\bibfnamefont {L.}~\bibnamefont {Cevey}}, \bibinfo
  {author} {\bibfnamefont {T.}~\bibnamefont {Wolf}}, \bibinfo {author}
  {\bibfnamefont {C.}~\bibnamefont {Meingast}}, \ and\ \bibinfo {author}
  {\bibfnamefont {R.}~\bibnamefont {Lortz}},\ }\href@noop {} {\bibfield
  {journal} {\bibinfo  {journal} {Phys. Rev. B}\ }\textbf {\bibinfo {volume}
  {87}},\ \bibinfo {pages} {214523} (\bibinfo {year} {2013})}\BibitemShut
  {NoStop}%
\bibitem [{\citenamefont {Koshelev}\ \emph {et~al.}(2019)\citenamefont
  {Koshelev}, \citenamefont {Willa}, \citenamefont {Willa}, \citenamefont
  {Smylie}, \citenamefont {Bao}, \citenamefont {Chung}, \citenamefont
  {Kanatzidis}, \citenamefont {Kwok},\ and\ \citenamefont {Welp}}]{Koshelev19}%
  \BibitemOpen
  \bibfield  {author} {\bibinfo {author} {\bibfnamefont {A.~E.}\ \bibnamefont
  {Koshelev}}, \bibinfo {author} {\bibfnamefont {K.}~\bibnamefont {Willa}},
  \bibinfo {author} {\bibfnamefont {R.}~\bibnamefont {Willa}}, \bibinfo
  {author} {\bibfnamefont {M.~P.}\ \bibnamefont {Smylie}}, \bibinfo {author}
  {\bibfnamefont {J.-K.}\ \bibnamefont {Bao}}, \bibinfo {author} {\bibfnamefont
  {D.~Y.}\ \bibnamefont {Chung}}, \bibinfo {author} {\bibfnamefont {M.~G.}\
  \bibnamefont {Kanatzidis}}, \bibinfo {author} {\bibfnamefont {W.-K.}\
  \bibnamefont {Kwok}}, \ and\ \bibinfo {author} {\bibfnamefont
  {U.}~\bibnamefont {Welp}},\ }\href@noop {} {\bibfield  {journal} {\bibinfo
  {journal} {Phys. Rev. B}\ }\textbf {\bibinfo {volume} {100}},\ \bibinfo
  {pages} {094518} (\bibinfo {year} {2019})}\BibitemShut {NoStop}%
\bibitem [{\citenamefont {Blatter}\ and\ \citenamefont
  {Ivlev}(1993)}]{Blatter93}%
  \BibitemOpen
  \bibfield  {author} {\bibinfo {author} {\bibfnamefont {G.}~\bibnamefont
  {Blatter}}\ and\ \bibinfo {author} {\bibfnamefont {B.}~\bibnamefont
  {Ivlev}},\ }\href@noop {} {\bibfield  {journal} {\bibinfo  {journal} {Phys.
  Rev. Lett.}\ }\textbf {\bibinfo {volume} {70}},\ \bibinfo {pages} {2621}
  (\bibinfo {year} {1993})}\BibitemShut {NoStop}%
\bibitem [{\citenamefont {Blatter}\ \emph
  {et~al.}(1994{\natexlab{b}})\citenamefont {Blatter}, \citenamefont {Ivlev},
  \citenamefont {Kagan}, \citenamefont {Theunissen}, \citenamefont
  {Volokitin},\ and\ \citenamefont {Kes}}]{Blatter94-1}%
  \BibitemOpen
  \bibfield  {author} {\bibinfo {author} {\bibfnamefont {G.}~\bibnamefont
  {Blatter}}, \bibinfo {author} {\bibfnamefont {B.}~\bibnamefont {Ivlev}},
  \bibinfo {author} {\bibfnamefont {Y.}~\bibnamefont {Kagan}}, \bibinfo
  {author} {\bibfnamefont {M.}~\bibnamefont {Theunissen}}, \bibinfo {author}
  {\bibfnamefont {Y.}~\bibnamefont {Volokitin}}, \ and\ \bibinfo {author}
  {\bibfnamefont {P.}~\bibnamefont {Kes}},\ }\href@noop {} {\bibfield
  {journal} {\bibinfo  {journal} {Phys. Rev. B}\ }\textbf {\bibinfo {volume}
  {50}},\ \bibinfo {pages} {13013} (\bibinfo {year}
  {1994}{\natexlab{b}})}\BibitemShut {NoStop}%
\bibitem [{\citenamefont {Blatter}\ and\ \citenamefont
  {Ivlev}(1994)}]{Blatter94-2}%
  \BibitemOpen
  \bibfield  {author} {\bibinfo {author} {\bibfnamefont {G.}~\bibnamefont
  {Blatter}}\ and\ \bibinfo {author} {\bibfnamefont {B.~I.}\ \bibnamefont
  {Ivlev}},\ }\href@noop {} {\bibfield  {journal} {\bibinfo  {journal} {Phys.
  Rev. B}\ }\textbf {\bibinfo {volume} {50}},\ \bibinfo {pages} {10272}
  (\bibinfo {year} {1994})}\BibitemShut {NoStop}%
\bibitem [{\citenamefont {Adachi}\ and\ \citenamefont
  {Ikeda}(2003)}]{Adachi03}%
  \BibitemOpen
  \bibfield  {author} {\bibinfo {author} {\bibfnamefont {H.}~\bibnamefont
  {Adachi}}\ and\ \bibinfo {author} {\bibfnamefont {R.}~\bibnamefont {Ikeda}},\
  }\href@noop {} {\bibfield  {journal} {\bibinfo  {journal} {Phys. Rev. B}\
  }\textbf {\bibinfo {volume} {68}},\ \bibinfo {pages} {184510} (\bibinfo
  {year} {2003})}\BibitemShut {NoStop}%
\bibitem [{\citenamefont {Lortz}\ \emph
  {et~al.}(2007{\natexlab{a}})\citenamefont {Lortz}, \citenamefont {Wang},
  \citenamefont {Demuer}, \citenamefont {B\"ottger}, \citenamefont {Bergk},
  \citenamefont {Zwicknagl}, \citenamefont {Nakazawa},\ and\ \citenamefont
  {Wosnitza}}]{Lortz07fflo}%
  \BibitemOpen
  \bibfield  {author} {\bibinfo {author} {\bibfnamefont {R.}~\bibnamefont
  {Lortz}}, \bibinfo {author} {\bibfnamefont {Y.}~\bibnamefont {Wang}},
  \bibinfo {author} {\bibfnamefont {A.}~\bibnamefont {Demuer}}, \bibinfo
  {author} {\bibfnamefont {P.~H.~M.}\ \bibnamefont {B\"ottger}}, \bibinfo
  {author} {\bibfnamefont {B.}~\bibnamefont {Bergk}}, \bibinfo {author}
  {\bibfnamefont {G.}~\bibnamefont {Zwicknagl}}, \bibinfo {author}
  {\bibfnamefont {Y.}~\bibnamefont {Nakazawa}}, \ and\ \bibinfo {author}
  {\bibfnamefont {J.}~\bibnamefont {Wosnitza}},\ }\href@noop {} {\bibfield
  {journal} {\bibinfo  {journal} {Phys. Rev. Lett.}\ }\textbf {\bibinfo
  {volume} {99}},\ \bibinfo {pages} {187002} (\bibinfo {year}
  {2007}{\natexlab{a}})}\BibitemShut {NoStop}%
\bibitem [{\citenamefont {Bianchi}\ \emph {et~al.}(2003)\citenamefont
  {Bianchi}, \citenamefont {Movshovich}, \citenamefont {Capan}, \citenamefont
  {Pagliuso},\ and\ \citenamefont {Sarrao}}]{Bianchi03}%
  \BibitemOpen
  \bibfield  {author} {\bibinfo {author} {\bibfnamefont {A.}~\bibnamefont
  {Bianchi}}, \bibinfo {author} {\bibfnamefont {R.}~\bibnamefont {Movshovich}},
  \bibinfo {author} {\bibfnamefont {C.}~\bibnamefont {Capan}}, \bibinfo
  {author} {\bibfnamefont {P.~G.}\ \bibnamefont {Pagliuso}}, \ and\ \bibinfo
  {author} {\bibfnamefont {J.~L.}\ \bibnamefont {Sarrao}},\ }\href@noop {}
  {\bibfield  {journal} {\bibinfo  {journal} {Phys. Rev. Lett.}\ }\textbf
  {\bibinfo {volume} {91}},\ \bibinfo {pages} {187004} (\bibinfo {year}
  {2003})}\BibitemShut {NoStop}%
\bibitem [{\citenamefont {Matsuda}\ and\ \citenamefont
  {Shimahara}(2007)}]{Matsuda07}%
  \BibitemOpen
  \bibfield  {author} {\bibinfo {author} {\bibfnamefont {Y.}~\bibnamefont
  {Matsuda}}\ and\ \bibinfo {author} {\bibfnamefont {H.}~\bibnamefont
  {Shimahara}},\ }\href@noop {} {\bibfield  {journal} {\bibinfo  {journal} {J.
  Phys. Soc. Jpn}\ }\textbf {\bibinfo {volume} {76}},\ \bibinfo {pages}
  {051005} (\bibinfo {year} {2007})}\BibitemShut {NoStop}%
\bibitem [{\citenamefont {Fulde}\ and\ \citenamefont
  {Ferrell}(1964)}]{Fulde64}%
  \BibitemOpen
  \bibfield  {author} {\bibinfo {author} {\bibfnamefont {P.}~\bibnamefont
  {Fulde}}\ and\ \bibinfo {author} {\bibfnamefont {R.~A.}\ \bibnamefont
  {Ferrell}},\ }\href@noop {} {\bibfield  {journal} {\bibinfo  {journal} {Phys.
  Rev.}\ }\textbf {\bibinfo {volume} {135}},\ \bibinfo {pages} {A550} (\bibinfo
  {year} {1964})}\BibitemShut {NoStop}%
\bibitem [{\citenamefont {Larkin}\ and\ \citenamefont {N.}(1965)}]{Larkin65}%
  \BibitemOpen
  \bibfield  {author} {\bibinfo {author} {\bibfnamefont {A.~I.}\ \bibnamefont
  {Larkin}}\ and\ \bibinfo {author} {\bibfnamefont {O.~Y.}\ \bibnamefont
  {N.}},\ }\href@noop {} {\bibfield  {journal} {\bibinfo  {journal} {Sov. Phys.
  JETP}\ }\textbf {\bibinfo {volume} {20}},\ \bibinfo {pages} {762} (\bibinfo
  {year} {1965})}\BibitemShut {NoStop}%
\bibitem [{\citenamefont {Lortz}\ \emph
  {et~al.}(2007{\natexlab{b}})\citenamefont {Lortz}, \citenamefont {Meingast},
  \citenamefont {Rykov},\ and\ \citenamefont {Tajima}}]{Lortz07}%
  \BibitemOpen
  \bibfield  {author} {\bibinfo {author} {\bibfnamefont {R.}~\bibnamefont
  {Lortz}}, \bibinfo {author} {\bibfnamefont {C.}~\bibnamefont {Meingast}},
  \bibinfo {author} {\bibfnamefont {A.~I.}\ \bibnamefont {Rykov}}, \ and\
  \bibinfo {author} {\bibfnamefont {S.}~\bibnamefont {Tajima}},\ }\href@noop {}
  {\bibfield  {journal} {\bibinfo  {journal} {J. Low Temp. Phys.}\ }\textbf
  {\bibinfo {volume} {147}},\ \bibinfo {pages} {365} (\bibinfo {year}
  {2007}{\natexlab{b}})}\BibitemShut {NoStop}%
\bibitem [{\citenamefont {Wosnitza}(2018)}]{Wosnitza18}%
  \BibitemOpen
  \bibfield  {author} {\bibinfo {author} {\bibfnamefont {J.}~\bibnamefont
  {Wosnitza}},\ }\href@noop {} {\bibfield  {journal} {\bibinfo  {journal} {Ann.
  Phys. (Berlin)}\ }\textbf {\bibinfo {volume} {530}},\ \bibinfo {pages}
  {1700282} (\bibinfo {year} {2018})}\BibitemShut {NoStop}%
\bibitem [{\citenamefont {Kenzelmann}\ \emph {et~al.}(2008)\citenamefont
  {Kenzelmann}, \citenamefont {Strässle}, \citenamefont {Niedermayer},
  \citenamefont {Sigrist}, \citenamefont {Padmanabhan}, \citenamefont
  {Zolliker}, \citenamefont {Bianchi}, \citenamefont {Movshovich},
  \citenamefont {Bauer}, \citenamefont {Sarrao},\ and\ \citenamefont
  {Thompson}}]{Kenzelmann08}%
  \BibitemOpen
  \bibfield  {author} {\bibinfo {author} {\bibfnamefont {M.}~\bibnamefont
  {Kenzelmann}}, \bibinfo {author} {\bibfnamefont {T.}~\bibnamefont
  {Strässle}}, \bibinfo {author} {\bibfnamefont {C.}~\bibnamefont
  {Niedermayer}}, \bibinfo {author} {\bibfnamefont {M.}~\bibnamefont
  {Sigrist}}, \bibinfo {author} {\bibfnamefont {B.}~\bibnamefont
  {Padmanabhan}}, \bibinfo {author} {\bibfnamefont {M.}~\bibnamefont
  {Zolliker}}, \bibinfo {author} {\bibfnamefont {A.~D.}\ \bibnamefont
  {Bianchi}}, \bibinfo {author} {\bibfnamefont {R.}~\bibnamefont {Movshovich}},
  \bibinfo {author} {\bibfnamefont {E.~D.}\ \bibnamefont {Bauer}}, \bibinfo
  {author} {\bibfnamefont {J.~L.}\ \bibnamefont {Sarrao}}, \ and\ \bibinfo
  {author} {\bibfnamefont {J.~D.}\ \bibnamefont {Thompson}},\ }\href@noop {}
  {\bibfield  {journal} {\bibinfo  {journal} {Science}\ }\textbf {\bibinfo
  {volume} {321}},\ \bibinfo {pages} {1652} (\bibinfo {year}
  {2008})}\BibitemShut {NoStop}%
\bibitem [{\citenamefont {Kenzelmann}\ \emph {et~al.}(2010)\citenamefont
  {Kenzelmann}, \citenamefont {Gerber}, \citenamefont {Egetenmeyer},
  \citenamefont {Gavilano}, \citenamefont {Str\"assle}, \citenamefont
  {Bianchi}, \citenamefont {Ressouche}, \citenamefont {Movshovich},
  \citenamefont {Bauer}, \citenamefont {Sarrao},\ and\ \citenamefont
  {Thompson}}]{Kenzelmann10}%
  \BibitemOpen
  \bibfield  {author} {\bibinfo {author} {\bibfnamefont {M.}~\bibnamefont
  {Kenzelmann}}, \bibinfo {author} {\bibfnamefont {S.}~\bibnamefont {Gerber}},
  \bibinfo {author} {\bibfnamefont {N.}~\bibnamefont {Egetenmeyer}}, \bibinfo
  {author} {\bibfnamefont {J.~L.}\ \bibnamefont {Gavilano}}, \bibinfo {author}
  {\bibfnamefont {T.}~\bibnamefont {Str\"assle}}, \bibinfo {author}
  {\bibfnamefont {A.~D.}\ \bibnamefont {Bianchi}}, \bibinfo {author}
  {\bibfnamefont {E.}~\bibnamefont {Ressouche}}, \bibinfo {author}
  {\bibfnamefont {R.}~\bibnamefont {Movshovich}}, \bibinfo {author}
  {\bibfnamefont {E.~D.}\ \bibnamefont {Bauer}}, \bibinfo {author}
  {\bibfnamefont {J.~L.}\ \bibnamefont {Sarrao}}, \ and\ \bibinfo {author}
  {\bibfnamefont {J.~D.}\ \bibnamefont {Thompson}},\ }\href@noop {} {\bibfield
  {journal} {\bibinfo  {journal} {Phys. Rev. Lett.}\ }\textbf {\bibinfo
  {volume} {104}},\ \bibinfo {pages} {127001} (\bibinfo {year}
  {2010})}\BibitemShut {NoStop}%
\bibitem [{\citenamefont {Kumagai}\ \emph {et~al.}(2011)\citenamefont
  {Kumagai}, \citenamefont {Shishido}, \citenamefont {Shibauchi},\ and\
  \citenamefont {Matsuda}}]{Kumagai11}%
  \BibitemOpen
  \bibfield  {author} {\bibinfo {author} {\bibfnamefont {K.}~\bibnamefont
  {Kumagai}}, \bibinfo {author} {\bibfnamefont {H.}~\bibnamefont {Shishido}},
  \bibinfo {author} {\bibfnamefont {T.}~\bibnamefont {Shibauchi}}, \ and\
  \bibinfo {author} {\bibfnamefont {Y.}~\bibnamefont {Matsuda}},\ }\href@noop
  {} {\bibfield  {journal} {\bibinfo  {journal} {Phys. Rev. Lett.}\ }\textbf
  {\bibinfo {volume} {106}},\ \bibinfo {pages} {137004} (\bibinfo {year}
  {2011})}\BibitemShut {NoStop}%
\bibitem [{\citenamefont {Murray}\ and\ \citenamefont {Te\ifmmode
  \check{s}\else \v{s}\fi{}anovi\ifmmode~\acute{c}\else
  \'{c}\fi{}}(2010)}]{Murray10}%
  \BibitemOpen
  \bibfield  {author} {\bibinfo {author} {\bibfnamefont {J.~M.}\ \bibnamefont
  {Murray}}\ and\ \bibinfo {author} {\bibfnamefont {Z.}~\bibnamefont
  {Te\ifmmode \check{s}\else \v{s}\fi{}anovi\ifmmode~\acute{c}\else
  \'{c}\fi{}}},\ }\href@noop {} {\bibfield  {journal} {\bibinfo  {journal}
  {Phys. Rev. Lett.}\ }\textbf {\bibinfo {volume} {105}},\ \bibinfo {pages}
  {037006} (\bibinfo {year} {2010})}\BibitemShut {NoStop}%
\bibitem [{\citenamefont {Hou}\ \emph {et~al.}(2015)\citenamefont {Hou},
  \citenamefont {Burger}, \citenamefont {Mak}, \citenamefont {Hardy},
  \citenamefont {Wolf}, \citenamefont {Meingast},\ and\ \citenamefont
  {Lortz}}]{Hou15}%
  \BibitemOpen
  \bibfield  {author} {\bibinfo {author} {\bibfnamefont {J.}~\bibnamefont
  {Hou}}, \bibinfo {author} {\bibfnamefont {P.}~\bibnamefont {Burger}},
  \bibinfo {author} {\bibfnamefont {H.~K.}\ \bibnamefont {Mak}}, \bibinfo
  {author} {\bibfnamefont {F.}~\bibnamefont {Hardy}}, \bibinfo {author}
  {\bibfnamefont {T.}~\bibnamefont {Wolf}}, \bibinfo {author} {\bibfnamefont
  {C.}~\bibnamefont {Meingast}}, \ and\ \bibinfo {author} {\bibfnamefont
  {R.}~\bibnamefont {Lortz}},\ }\href@noop {} {\bibfield  {journal} {\bibinfo
  {journal} {Phys. Rev. B}\ }\textbf {\bibinfo {volume} {92}},\ \bibinfo
  {pages} {064502} (\bibinfo {year} {2015})}\BibitemShut {NoStop}%
\bibitem [{\citenamefont {Burger}\ \emph {et~al.}(2013)\citenamefont {Burger},
  \citenamefont {Hardy}, \citenamefont {Aoki}, \citenamefont {B\"ohmer},
  \citenamefont {Eder}, \citenamefont {Heid}, \citenamefont {Wolf},
  \citenamefont {Schweiss}, \citenamefont {Fromknecht}, \citenamefont
  {Jackson}, \citenamefont {Paulsen},\ and\ \citenamefont
  {Meingast}}]{Burger13}%
  \BibitemOpen
  \bibfield  {author} {\bibinfo {author} {\bibfnamefont {P.}~\bibnamefont
  {Burger}}, \bibinfo {author} {\bibfnamefont {F.}~\bibnamefont {Hardy}},
  \bibinfo {author} {\bibfnamefont {D.}~\bibnamefont {Aoki}}, \bibinfo {author}
  {\bibfnamefont {A.~E.}\ \bibnamefont {B\"ohmer}}, \bibinfo {author}
  {\bibfnamefont {R.}~\bibnamefont {Eder}}, \bibinfo {author} {\bibfnamefont
  {R.}~\bibnamefont {Heid}}, \bibinfo {author} {\bibfnamefont {T.}~\bibnamefont
  {Wolf}}, \bibinfo {author} {\bibfnamefont {P.}~\bibnamefont {Schweiss}},
  \bibinfo {author} {\bibfnamefont {R.}~\bibnamefont {Fromknecht}}, \bibinfo
  {author} {\bibfnamefont {M.~J.}\ \bibnamefont {Jackson}}, \bibinfo {author}
  {\bibfnamefont {C.}~\bibnamefont {Paulsen}}, \ and\ \bibinfo {author}
  {\bibfnamefont {C.}~\bibnamefont {Meingast}},\ }\href@noop {} {\bibfield
  {journal} {\bibinfo  {journal} {Phys. Rev. B}\ }\textbf {\bibinfo {volume}
  {88}},\ \bibinfo {pages} {014517} (\bibinfo {year} {2013})}\BibitemShut
  {NoStop}%
\bibitem [{\citenamefont {Zocco}\ \emph {et~al.}(2013)\citenamefont {Zocco},
  \citenamefont {Grube}, \citenamefont {Eilers}, \citenamefont {Wolf},\ and\
  \citenamefont {L\"ohneysen}}]{Zocco13}%
  \BibitemOpen
  \bibfield  {author} {\bibinfo {author} {\bibfnamefont {D.~A.}\ \bibnamefont
  {Zocco}}, \bibinfo {author} {\bibfnamefont {K.}~\bibnamefont {Grube}},
  \bibinfo {author} {\bibfnamefont {F.}~\bibnamefont {Eilers}}, \bibinfo
  {author} {\bibfnamefont {T.}~\bibnamefont {Wolf}}, \ and\ \bibinfo {author}
  {\bibfnamefont {H.~v.}\ \bibnamefont {L\"ohneysen}},\ }\href@noop {}
  {\bibfield  {journal} {\bibinfo  {journal} {Phys. Rev. Lett.}\ }\textbf
  {\bibinfo {volume} {111}},\ \bibinfo {pages} {057007} (\bibinfo {year}
  {2013})}\BibitemShut {NoStop}%
\bibitem [{\citenamefont {Cho}\ \emph {et~al.}(2017)\citenamefont {Cho},
  \citenamefont {Yang}, \citenamefont {Yuan}, \citenamefont {Shen},
  \citenamefont {Wolf},\ and\ \citenamefont {Lortz}}]{Cho17}%
  \BibitemOpen
  \bibfield  {author} {\bibinfo {author} {\bibfnamefont {C.-w.}\ \bibnamefont
  {Cho}}, \bibinfo {author} {\bibfnamefont {J.~H.}\ \bibnamefont {Yang}},
  \bibinfo {author} {\bibfnamefont {N.~F.~Q.}\ \bibnamefont {Yuan}}, \bibinfo
  {author} {\bibfnamefont {J.}~\bibnamefont {Shen}}, \bibinfo {author}
  {\bibfnamefont {T.}~\bibnamefont {Wolf}}, \ and\ \bibinfo {author}
  {\bibfnamefont {R.}~\bibnamefont {Lortz}},\ }\href@noop {} {\bibfield
  {journal} {\bibinfo  {journal} {Phys. Rev. Lett.}\ }\textbf {\bibinfo
  {volume} {119}},\ \bibinfo {pages} {217002} (\bibinfo {year}
  {2017})}\BibitemShut {NoStop}%
\bibitem [{\citenamefont {Böhmer}\ and\ \citenamefont
  {Kreisel}(2017)}]{Boehmer17}%
  \BibitemOpen
  \bibfield  {author} {\bibinfo {author} {\bibfnamefont {A.~E.}\ \bibnamefont
  {Böhmer}}\ and\ \bibinfo {author} {\bibfnamefont {A.}~\bibnamefont
  {Kreisel}},\ }\href@noop {} {\bibfield  {journal} {\bibinfo  {journal} {J.
  Phys.: Condens. Matter}\ }\textbf {\bibinfo {volume} {30}},\ \bibinfo {pages}
  {023001} (\bibinfo {year} {2017})}\BibitemShut {NoStop}%
\bibitem [{\citenamefont {Coldea}\ and\ \citenamefont
  {Watson}(2018)}]{Coldea18}%
  \BibitemOpen
  \bibfield  {author} {\bibinfo {author} {\bibfnamefont {A.}~\bibnamefont
  {Coldea}}\ and\ \bibinfo {author} {\bibfnamefont {M.~D.}\ \bibnamefont
  {Watson}},\ }\href@noop {} {\bibfield  {journal} {\bibinfo  {journal} {Annu.
  Rev. Condens. Matter Phys.}\ }\textbf {\bibinfo {volume} {9}},\ \bibinfo
  {pages} {125} (\bibinfo {year} {2018})}\BibitemShut {NoStop}%
\bibitem [{\citenamefont {Sprau}\ \emph {et~al.}(2017)\citenamefont {Sprau},
  \citenamefont {Kostin}, \citenamefont {Kreisel}, \citenamefont {Böhmer},
  \citenamefont {Taufour}, \citenamefont {Canfield}, \citenamefont {Mukherjee},
  \citenamefont {Hirschfeld}, \citenamefont {Andersen},\ and\ \citenamefont
  {Davis}}]{Sprau17}%
  \BibitemOpen
  \bibfield  {author} {\bibinfo {author} {\bibfnamefont {P.~O.}\ \bibnamefont
  {Sprau}}, \bibinfo {author} {\bibfnamefont {A.}~\bibnamefont {Kostin}},
  \bibinfo {author} {\bibfnamefont {A.}~\bibnamefont {Kreisel}}, \bibinfo
  {author} {\bibfnamefont {A.~E.}\ \bibnamefont {Böhmer}}, \bibinfo {author}
  {\bibfnamefont {V.}~\bibnamefont {Taufour}}, \bibinfo {author} {\bibfnamefont
  {P.~C.}\ \bibnamefont {Canfield}}, \bibinfo {author} {\bibfnamefont
  {S.}~\bibnamefont {Mukherjee}}, \bibinfo {author} {\bibfnamefont {P.~J.}\
  \bibnamefont {Hirschfeld}}, \bibinfo {author} {\bibfnamefont {B.~M.}\
  \bibnamefont {Andersen}}, \ and\ \bibinfo {author} {\bibfnamefont {J.~C.~S.}\
  \bibnamefont {Davis}},\ }\href@noop {} {\bibfield  {journal} {\bibinfo
  {journal} {Science}\ }\textbf {\bibinfo {volume} {357}},\ \bibinfo {pages}
  {75} (\bibinfo {year} {2017})}\BibitemShut {NoStop}%
\bibitem [{\citenamefont {Rhodes}\ \emph {et~al.}(2018)\citenamefont {Rhodes},
  \citenamefont {Watson}, \citenamefont {Haghighirad}, \citenamefont
  {Evtushinsky}, \citenamefont {Eschrig},\ and\ \citenamefont
  {Kim}}]{Rhodes18}%
  \BibitemOpen
  \bibfield  {author} {\bibinfo {author} {\bibfnamefont {L.~C.}\ \bibnamefont
  {Rhodes}}, \bibinfo {author} {\bibfnamefont {M.~D.}\ \bibnamefont {Watson}},
  \bibinfo {author} {\bibfnamefont {A.~A.}\ \bibnamefont {Haghighirad}},
  \bibinfo {author} {\bibfnamefont {D.~V.}\ \bibnamefont {Evtushinsky}},
  \bibinfo {author} {\bibfnamefont {M.}~\bibnamefont {Eschrig}}, \ and\
  \bibinfo {author} {\bibfnamefont {T.~K.}\ \bibnamefont {Kim}},\ }\href@noop
  {} {\bibfield  {journal} {\bibinfo  {journal} {Phys. Rev. B}\ }\textbf
  {\bibinfo {volume} {98}},\ \bibinfo {pages} {180503} (\bibinfo {year}
  {2018})}\BibitemShut {NoStop}%
\bibitem [{\citenamefont {Kreisel}\ \emph {et~al.}(2017)\citenamefont
  {Kreisel}, \citenamefont {Andersen}, \citenamefont {Sprau}, \citenamefont
  {Kostin}, \citenamefont {Davis},\ and\ \citenamefont
  {Hirschfeld}}]{Kreisel17}%
  \BibitemOpen
  \bibfield  {author} {\bibinfo {author} {\bibfnamefont {A.}~\bibnamefont
  {Kreisel}}, \bibinfo {author} {\bibfnamefont {B.~M.}\ \bibnamefont
  {Andersen}}, \bibinfo {author} {\bibfnamefont {P.~O.}\ \bibnamefont {Sprau}},
  \bibinfo {author} {\bibfnamefont {A.}~\bibnamefont {Kostin}}, \bibinfo
  {author} {\bibfnamefont {J.~C.~S.}\ \bibnamefont {Davis}}, \ and\ \bibinfo
  {author} {\bibfnamefont {P.~J.}\ \bibnamefont {Hirschfeld}},\ }\href@noop {}
  {\bibfield  {journal} {\bibinfo  {journal} {Phys. Rev. B}\ }\textbf {\bibinfo
  {volume} {95}},\ \bibinfo {pages} {174504} (\bibinfo {year}
  {2017})}\BibitemShut {NoStop}%
\bibitem [{\citenamefont {Kang}\ \emph {et~al.}(2018)\citenamefont {Kang},
  \citenamefont {Fernandes},\ and\ \citenamefont {Chubukov}}]{Kang18}%
  \BibitemOpen
  \bibfield  {author} {\bibinfo {author} {\bibfnamefont {J.}~\bibnamefont
  {Kang}}, \bibinfo {author} {\bibfnamefont {R.~M.}\ \bibnamefont {Fernandes}},
  \ and\ \bibinfo {author} {\bibfnamefont {A.}~\bibnamefont {Chubukov}},\
  }\href@noop {} {\bibfield  {journal} {\bibinfo  {journal} {Phys. Rev. Lett.}\
  }\textbf {\bibinfo {volume} {120}},\ \bibinfo {pages} {267001} (\bibinfo
  {year} {2018})}\BibitemShut {NoStop}%
\bibitem [{\citenamefont {Benfatto}\ \emph {et~al.}(2018)\citenamefont
  {Benfatto}, \citenamefont {Valenzuela},\ and\ \citenamefont
  {Fanfarillo}}]{Benfatto18}%
  \BibitemOpen
  \bibfield  {author} {\bibinfo {author} {\bibfnamefont {L.}~\bibnamefont
  {Benfatto}}, \bibinfo {author} {\bibfnamefont {B.}~\bibnamefont
  {Valenzuela}}, \ and\ \bibinfo {author} {\bibfnamefont {L.}~\bibnamefont
  {Fanfarillo}},\ }\href@noop {} {\bibfield  {journal} {\bibinfo  {journal}
  {npj Quantum Mater.}\ }\textbf {\bibinfo {volume} {3}},\ \bibinfo {pages}
  {56} (\bibinfo {year} {2018})}\BibitemShut {NoStop}%
\bibitem [{\citenamefont {Cercellier}\ \emph {et~al.}(2019)\citenamefont
  {Cercellier}, \citenamefont {Rodi\`ere}, \citenamefont {Toulemonde},
  \citenamefont {Marcenat},\ and\ \citenamefont {Klein}}]{Cercellier19}%
  \BibitemOpen
  \bibfield  {author} {\bibinfo {author} {\bibfnamefont {H.}~\bibnamefont
  {Cercellier}}, \bibinfo {author} {\bibfnamefont {P.}~\bibnamefont
  {Rodi\`ere}}, \bibinfo {author} {\bibfnamefont {P.}~\bibnamefont
  {Toulemonde}}, \bibinfo {author} {\bibfnamefont {C.}~\bibnamefont
  {Marcenat}}, \ and\ \bibinfo {author} {\bibfnamefont {T.}~\bibnamefont
  {Klein}},\ }\href@noop {} {\bibfield  {journal} {\bibinfo  {journal} {Phys.
  Rev. B}\ }\textbf {\bibinfo {volume} {100}},\ \bibinfo {pages} {104516}
  (\bibinfo {year} {2019})}\BibitemShut {NoStop}%
\bibitem [{\citenamefont {Terashima}\ \emph {et~al.}(2014)\citenamefont
  {Terashima}, \citenamefont {Kikugawa}, \citenamefont {Kiswandhi},
  \citenamefont {Choi}, \citenamefont {Brooks}, \citenamefont {Kasahara},
  \citenamefont {Watashige}, \citenamefont {Ikeda}, \citenamefont {Shibauchi},
  \citenamefont {Matsuda}, \citenamefont {Wolf}, \citenamefont {B\"ohmer},
  \citenamefont {Hardy}, \citenamefont {Meingast}, \citenamefont {L\"ohneysen},
  \citenamefont {Suzuki}, \citenamefont {Arita},\ and\ \citenamefont
  {Uji}}]{Terashima14}%
  \BibitemOpen
  \bibfield  {author} {\bibinfo {author} {\bibfnamefont {T.}~\bibnamefont
  {Terashima}}, \bibinfo {author} {\bibfnamefont {N.}~\bibnamefont {Kikugawa}},
  \bibinfo {author} {\bibfnamefont {A.}~\bibnamefont {Kiswandhi}}, \bibinfo
  {author} {\bibfnamefont {E.-S.}\ \bibnamefont {Choi}}, \bibinfo {author}
  {\bibfnamefont {J.~S.}\ \bibnamefont {Brooks}}, \bibinfo {author}
  {\bibfnamefont {S.}~\bibnamefont {Kasahara}}, \bibinfo {author}
  {\bibfnamefont {T.}~\bibnamefont {Watashige}}, \bibinfo {author}
  {\bibfnamefont {H.}~\bibnamefont {Ikeda}}, \bibinfo {author} {\bibfnamefont
  {T.}~\bibnamefont {Shibauchi}}, \bibinfo {author} {\bibfnamefont
  {Y.}~\bibnamefont {Matsuda}}, \bibinfo {author} {\bibfnamefont
  {T.}~\bibnamefont {Wolf}}, \bibinfo {author} {\bibfnamefont {A.~E.}\
  \bibnamefont {B\"ohmer}}, \bibinfo {author} {\bibfnamefont {F.}~\bibnamefont
  {Hardy}}, \bibinfo {author} {\bibfnamefont {C.}~\bibnamefont {Meingast}},
  \bibinfo {author} {\bibfnamefont {H.~v.}\ \bibnamefont {L\"ohneysen}},
  \bibinfo {author} {\bibfnamefont {M.-T.}\ \bibnamefont {Suzuki}}, \bibinfo
  {author} {\bibfnamefont {R.}~\bibnamefont {Arita}}, \ and\ \bibinfo {author}
  {\bibfnamefont {S.}~\bibnamefont {Uji}},\ }\href@noop {} {\bibfield
  {journal} {\bibinfo  {journal} {Phys. Rev. B}\ }\textbf {\bibinfo {volume}
  {90}},\ \bibinfo {pages} {144517} (\bibinfo {year} {2014})}\BibitemShut
  {NoStop}%
\bibitem [{\citenamefont {Yang}\ \emph {et~al.}(2017)\citenamefont {Yang},
  \citenamefont {Chen}, \citenamefont {Zhu}, \citenamefont {Xing},\ and\
  \citenamefont {Wen}}]{Yang17}%
  \BibitemOpen
  \bibfield  {author} {\bibinfo {author} {\bibfnamefont {H.}~\bibnamefont
  {Yang}}, \bibinfo {author} {\bibfnamefont {G.}~\bibnamefont {Chen}}, \bibinfo
  {author} {\bibfnamefont {X.}~\bibnamefont {Zhu}}, \bibinfo {author}
  {\bibfnamefont {J.}~\bibnamefont {Xing}}, \ and\ \bibinfo {author}
  {\bibfnamefont {H.-H.}\ \bibnamefont {Wen}},\ }\href@noop {} {\bibfield
  {journal} {\bibinfo  {journal} {Phys. Rev. B}\ }\textbf {\bibinfo {volume}
  {96}},\ \bibinfo {pages} {064501} (\bibinfo {year} {2017})}\BibitemShut
  {NoStop}%
\bibitem [{\citenamefont {Kasahara}\ \emph {et~al.}(2014)\citenamefont
  {Kasahara}, \citenamefont {Watashige}, \citenamefont {Hanaguri},
  \citenamefont {Kohsaka}, \citenamefont {Yamashita}, \citenamefont
  {Shimoyama}, \citenamefont {Mizukami}, \citenamefont {Endo}, \citenamefont
  {Ikeda}, \citenamefont {Aoyama}, \citenamefont {Terashima}, \citenamefont
  {Uji}, \citenamefont {Wolf}, \citenamefont {von Löhneysen}, \citenamefont
  {Shibauchi},\ and\ \citenamefont {Matsuda}}]{Kasahara14}%
  \BibitemOpen
  \bibfield  {author} {\bibinfo {author} {\bibfnamefont {S.}~\bibnamefont
  {Kasahara}}, \bibinfo {author} {\bibfnamefont {T.}~\bibnamefont {Watashige}},
  \bibinfo {author} {\bibfnamefont {T.}~\bibnamefont {Hanaguri}}, \bibinfo
  {author} {\bibfnamefont {Y.}~\bibnamefont {Kohsaka}}, \bibinfo {author}
  {\bibfnamefont {T.}~\bibnamefont {Yamashita}}, \bibinfo {author}
  {\bibfnamefont {Y.}~\bibnamefont {Shimoyama}}, \bibinfo {author}
  {\bibfnamefont {Y.}~\bibnamefont {Mizukami}}, \bibinfo {author}
  {\bibfnamefont {R.}~\bibnamefont {Endo}}, \bibinfo {author} {\bibfnamefont
  {H.}~\bibnamefont {Ikeda}}, \bibinfo {author} {\bibfnamefont
  {K.}~\bibnamefont {Aoyama}}, \bibinfo {author} {\bibfnamefont
  {T.}~\bibnamefont {Terashima}}, \bibinfo {author} {\bibfnamefont
  {S.}~\bibnamefont {Uji}}, \bibinfo {author} {\bibfnamefont {T.}~\bibnamefont
  {Wolf}}, \bibinfo {author} {\bibfnamefont {H.}~\bibnamefont {von
  Löhneysen}}, \bibinfo {author} {\bibfnamefont {T.}~\bibnamefont
  {Shibauchi}}, \ and\ \bibinfo {author} {\bibfnamefont {Y.}~\bibnamefont
  {Matsuda}},\ }\href@noop {} {\bibfield  {journal} {\bibinfo  {journal} {Proc.
  Natl. Acad. Sci. U.S.A.}\ }\textbf {\bibinfo {volume} {111}},\ \bibinfo
  {pages} {16309} (\bibinfo {year} {2014})}\BibitemShut {NoStop}%
\bibitem [{\citenamefont {Watashige}\ \emph {et~al.}(2017)\citenamefont
  {Watashige}, \citenamefont {Arsenijević}, \citenamefont {Yamashita},
  \citenamefont {Terazawa}, \citenamefont {Onishi}, \citenamefont {Opherden},
  \citenamefont {Kasahara}, \citenamefont {Tokiwa}, \citenamefont {Kasahara},
  \citenamefont {Shibauchi}, \citenamefont {von Löhneysen}, \citenamefont
  {Wosnitza},\ and\ \citenamefont {Matsuda}}]{Watashige17}%
  \BibitemOpen
  \bibfield  {author} {\bibinfo {author} {\bibfnamefont {T.}~\bibnamefont
  {Watashige}}, \bibinfo {author} {\bibfnamefont {S.}~\bibnamefont
  {Arsenijević}}, \bibinfo {author} {\bibfnamefont {T.}~\bibnamefont
  {Yamashita}}, \bibinfo {author} {\bibfnamefont {D.}~\bibnamefont {Terazawa}},
  \bibinfo {author} {\bibfnamefont {T.}~\bibnamefont {Onishi}}, \bibinfo
  {author} {\bibfnamefont {L.}~\bibnamefont {Opherden}}, \bibinfo {author}
  {\bibfnamefont {S.}~\bibnamefont {Kasahara}}, \bibinfo {author}
  {\bibfnamefont {Y.}~\bibnamefont {Tokiwa}}, \bibinfo {author} {\bibfnamefont
  {Y.}~\bibnamefont {Kasahara}}, \bibinfo {author} {\bibfnamefont
  {T.}~\bibnamefont {Shibauchi}}, \bibinfo {author} {\bibfnamefont
  {H.}~\bibnamefont {von Löhneysen}}, \bibinfo {author} {\bibfnamefont
  {J.}~\bibnamefont {Wosnitza}}, \ and\ \bibinfo {author} {\bibfnamefont
  {Y.}~\bibnamefont {Matsuda}},\ }\href@noop {} {\bibfield  {journal} {\bibinfo
   {journal} {J. Phys. Soc. Jpn}\ }\textbf {\bibinfo {volume} {86}},\ \bibinfo
  {pages} {014707} (\bibinfo {year} {2017})}\BibitemShut {NoStop}%
\bibitem [{\citenamefont {Kasahara}\ \emph {et~al.}(2019)\citenamefont
  {Kasahara}, \citenamefont {Sato}, \citenamefont {Licciardello}, \citenamefont
  {Čulo}, \citenamefont {Arsenijević}, \citenamefont {Ottenbros},
  \citenamefont {Tominaga}, \citenamefont {Böker}, \citenamefont {Eremin},
  \citenamefont {Shibauchi}, \citenamefont {Wosnitza}, \citenamefont {Hussey},\
  and\ \citenamefont {Matsuda}}]{Kasahara19}%
  \BibitemOpen
  \bibfield  {author} {\bibinfo {author} {\bibfnamefont {S.}~\bibnamefont
  {Kasahara}}, \bibinfo {author} {\bibfnamefont {Y.}~\bibnamefont {Sato}},
  \bibinfo {author} {\bibfnamefont {S.}~\bibnamefont {Licciardello}}, \bibinfo
  {author} {\bibfnamefont {M.}~\bibnamefont {Čulo}}, \bibinfo {author}
  {\bibfnamefont {S.}~\bibnamefont {Arsenijević}}, \bibinfo {author}
  {\bibfnamefont {T.}~\bibnamefont {Ottenbros}}, \bibinfo {author}
  {\bibfnamefont {T.}~\bibnamefont {Tominaga}}, \bibinfo {author}
  {\bibfnamefont {J.}~\bibnamefont {Böker}}, \bibinfo {author} {\bibfnamefont
  {I.}~\bibnamefont {Eremin}}, \bibinfo {author} {\bibfnamefont
  {T.}~\bibnamefont {Shibauchi}}, \bibinfo {author} {\bibfnamefont
  {J.}~\bibnamefont {Wosnitza}}, \bibinfo {author} {\bibfnamefont {N.~E.}\
  \bibnamefont {Hussey}}, \ and\ \bibinfo {author} {\bibfnamefont
  {Y.}~\bibnamefont {Matsuda}},\ }\href@noop {} {} (\bibinfo {year} {2019}),\
  \Eprint {http://arxiv.org/abs/1911.08237} {arXiv:1911.08237
  [cond-mat.supr-con]} \BibitemShut {NoStop}%
\bibitem [{\citenamefont {Brison}\ \emph {et~al.}(1997)\citenamefont {Brison},
  \citenamefont {Buzdin}, \citenamefont {Glémont}, \citenamefont {Thomas},\
  and\ \citenamefont {Flouquet}}]{Brison97}%
  \BibitemOpen
  \bibfield  {author} {\bibinfo {author} {\bibfnamefont {J.}~\bibnamefont
  {Brison}}, \bibinfo {author} {\bibfnamefont {A.}~\bibnamefont {Buzdin}},
  \bibinfo {author} {\bibfnamefont {L.}~\bibnamefont {Glémont}}, \bibinfo
  {author} {\bibfnamefont {F.}~\bibnamefont {Thomas}}, \ and\ \bibinfo {author}
  {\bibfnamefont {J.}~\bibnamefont {Flouquet}},\ }\href@noop {} {\bibfield
  {journal} {\bibinfo  {journal} {Physica B}\ }\textbf {\bibinfo {volume}
  {230}},\ \bibinfo {pages} {406} (\bibinfo {year} {1997})}\BibitemShut
  {NoStop}%
\bibitem [{\citenamefont {Hardy}\ \emph {et~al.}(2019)\citenamefont {Hardy},
  \citenamefont {He}, \citenamefont {Wang}, \citenamefont {Wolf}, \citenamefont
  {Schweiss}, \citenamefont {Merz}, \citenamefont {Barth}, \citenamefont
  {Adelmann}, \citenamefont {Eder}, \citenamefont {Haghighirad},\ and\
  \citenamefont {Meingast}}]{Hardy19}%
  \BibitemOpen
  \bibfield  {author} {\bibinfo {author} {\bibfnamefont {F.}~\bibnamefont
  {Hardy}}, \bibinfo {author} {\bibfnamefont {M.}~\bibnamefont {He}}, \bibinfo
  {author} {\bibfnamefont {L.}~\bibnamefont {Wang}}, \bibinfo {author}
  {\bibfnamefont {T.}~\bibnamefont {Wolf}}, \bibinfo {author} {\bibfnamefont
  {P.}~\bibnamefont {Schweiss}}, \bibinfo {author} {\bibfnamefont
  {M.}~\bibnamefont {Merz}}, \bibinfo {author} {\bibfnamefont {M.}~\bibnamefont
  {Barth}}, \bibinfo {author} {\bibfnamefont {P.}~\bibnamefont {Adelmann}},
  \bibinfo {author} {\bibfnamefont {R.}~\bibnamefont {Eder}}, \bibinfo {author}
  {\bibfnamefont {A.-A.}\ \bibnamefont {Haghighirad}}, \ and\ \bibinfo {author}
  {\bibfnamefont {C.}~\bibnamefont {Meingast}},\ }\href@noop {} {\bibfield
  {journal} {\bibinfo  {journal} {Phys. Rev. B}\ }\textbf {\bibinfo {volume}
  {99}},\ \bibinfo {pages} {035157} (\bibinfo {year} {2019})}\BibitemShut
  {NoStop}%
\bibitem [{\citenamefont {Michon}\ \emph {et~al.}(2019)\citenamefont {Michon},
  \citenamefont {Girod}, \citenamefont {Badoux}, \citenamefont {Kačmarčík},
  \citenamefont {Ma}, \citenamefont {Dragomir}, \citenamefont {Dabkowska},
  \citenamefont {Gaulin}, \citenamefont {Zhou}, \citenamefont {Pyon},
  \citenamefont {Takayama}, \citenamefont {Takagi}, \citenamefont {Verret},
  \citenamefont {Doiron-Leyraud}, \citenamefont {Marcenat}, \citenamefont
  {Taillefer},\ and\ \citenamefont {Klein}}]{Michon19}%
  \BibitemOpen
  \bibfield  {author} {\bibinfo {author} {\bibfnamefont {B.}~\bibnamefont
  {Michon}}, \bibinfo {author} {\bibfnamefont {C.}~\bibnamefont {Girod}},
  \bibinfo {author} {\bibfnamefont {S.}~\bibnamefont {Badoux}}, \bibinfo
  {author} {\bibfnamefont {J.}~\bibnamefont {Kačmarčík}}, \bibinfo {author}
  {\bibfnamefont {Q.}~\bibnamefont {Ma}}, \bibinfo {author} {\bibfnamefont
  {M.}~\bibnamefont {Dragomir}}, \bibinfo {author} {\bibfnamefont {H.~A.}\
  \bibnamefont {Dabkowska}}, \bibinfo {author} {\bibfnamefont {B.~D.}\
  \bibnamefont {Gaulin}}, \bibinfo {author} {\bibfnamefont {J.~S.}\
  \bibnamefont {Zhou}}, \bibinfo {author} {\bibfnamefont {S.}~\bibnamefont
  {Pyon}}, \bibinfo {author} {\bibfnamefont {T.}~\bibnamefont {Takayama}},
  \bibinfo {author} {\bibfnamefont {H.}~\bibnamefont {Takagi}}, \bibinfo
  {author} {\bibfnamefont {S.}~\bibnamefont {Verret}}, \bibinfo {author}
  {\bibfnamefont {N.}~\bibnamefont {Doiron-Leyraud}}, \bibinfo {author}
  {\bibfnamefont {C.}~\bibnamefont {Marcenat}}, \bibinfo {author}
  {\bibfnamefont {L.}~\bibnamefont {Taillefer}}, \ and\ \bibinfo {author}
  {\bibfnamefont {T.}~\bibnamefont {Klein}},\ }\href@noop {} {\bibfield
  {journal} {\bibinfo  {journal} {Nature}\ }\textbf {\bibinfo {volume} {567}},\
  \bibinfo {pages} {218} (\bibinfo {year} {2019})}\BibitemShut {NoStop}%
\bibitem [{\citenamefont {Meingast}\ \emph {et~al.}(1990)\citenamefont
  {Meingast}, \citenamefont {Blank}, \citenamefont {B\"urkle}, \citenamefont
  {Obst}, \citenamefont {Wolf}, \citenamefont {W\"uhl}, \citenamefont
  {Selvamanickam},\ and\ \citenamefont {Salama}}]{Meingast90}%
  \BibitemOpen
  \bibfield  {author} {\bibinfo {author} {\bibfnamefont {C.}~\bibnamefont
  {Meingast}}, \bibinfo {author} {\bibfnamefont {B.}~\bibnamefont {Blank}},
  \bibinfo {author} {\bibfnamefont {H.}~\bibnamefont {B\"urkle}}, \bibinfo
  {author} {\bibfnamefont {B.}~\bibnamefont {Obst}}, \bibinfo {author}
  {\bibfnamefont {T.}~\bibnamefont {Wolf}}, \bibinfo {author} {\bibfnamefont
  {H.}~\bibnamefont {W\"uhl}}, \bibinfo {author} {\bibfnamefont
  {V.}~\bibnamefont {Selvamanickam}}, \ and\ \bibinfo {author} {\bibfnamefont
  {K.}~\bibnamefont {Salama}},\ }\href@noop {} {\bibfield  {journal} {\bibinfo
  {journal} {Phys. Rev. B}\ }\textbf {\bibinfo {volume} {41}},\ \bibinfo
  {pages} {11299} (\bibinfo {year} {1990})}\BibitemShut {NoStop}%
\bibitem [{\citenamefont {Sun}\ \emph {et~al.}(2017)\citenamefont {Sun},
  \citenamefont {Kittaka}, \citenamefont {Nakamura}, \citenamefont
  {Sakakibara}, \citenamefont {Irie}, \citenamefont {Nomoto}, \citenamefont
  {Machida}, \citenamefont {Chen},\ and\ \citenamefont {Tamegai}}]{Sun17}%
  \BibitemOpen
  \bibfield  {author} {\bibinfo {author} {\bibfnamefont {Y.}~\bibnamefont
  {Sun}}, \bibinfo {author} {\bibfnamefont {S.}~\bibnamefont {Kittaka}},
  \bibinfo {author} {\bibfnamefont {S.}~\bibnamefont {Nakamura}}, \bibinfo
  {author} {\bibfnamefont {T.}~\bibnamefont {Sakakibara}}, \bibinfo {author}
  {\bibfnamefont {K.}~\bibnamefont {Irie}}, \bibinfo {author} {\bibfnamefont
  {T.}~\bibnamefont {Nomoto}}, \bibinfo {author} {\bibfnamefont
  {K.}~\bibnamefont {Machida}}, \bibinfo {author} {\bibfnamefont
  {J.}~\bibnamefont {Chen}}, \ and\ \bibinfo {author} {\bibfnamefont
  {T.}~\bibnamefont {Tamegai}},\ }\href@noop {} {\bibfield  {journal} {\bibinfo
   {journal} {Phys. Rev. B}\ }\textbf {\bibinfo {volume} {96}},\ \bibinfo
  {pages} {220505} (\bibinfo {year} {2017})}\BibitemShut {NoStop}%
\bibitem [{\citenamefont {Watashige}\ \emph {et~al.}(2015)\citenamefont
  {Watashige}, \citenamefont {Tsutsumi}, \citenamefont {Hanaguri},
  \citenamefont {Kohsaka}, \citenamefont {Kasahara}, \citenamefont {Furusaki},
  \citenamefont {Sigrist}, \citenamefont {Meingast}, \citenamefont {Wolf},
  \citenamefont {L\"ohneysen}, \citenamefont {Shibauchi},\ and\ \citenamefont
  {Matsuda}}]{Watashige15}%
  \BibitemOpen
  \bibfield  {author} {\bibinfo {author} {\bibfnamefont {T.}~\bibnamefont
  {Watashige}}, \bibinfo {author} {\bibfnamefont {Y.}~\bibnamefont {Tsutsumi}},
  \bibinfo {author} {\bibfnamefont {T.}~\bibnamefont {Hanaguri}}, \bibinfo
  {author} {\bibfnamefont {Y.}~\bibnamefont {Kohsaka}}, \bibinfo {author}
  {\bibfnamefont {S.}~\bibnamefont {Kasahara}}, \bibinfo {author}
  {\bibfnamefont {A.}~\bibnamefont {Furusaki}}, \bibinfo {author}
  {\bibfnamefont {M.}~\bibnamefont {Sigrist}}, \bibinfo {author} {\bibfnamefont
  {C.}~\bibnamefont {Meingast}}, \bibinfo {author} {\bibfnamefont
  {T.}~\bibnamefont {Wolf}}, \bibinfo {author} {\bibfnamefont {H.~v.}\
  \bibnamefont {L\"ohneysen}}, \bibinfo {author} {\bibfnamefont
  {T.}~\bibnamefont {Shibauchi}}, \ and\ \bibinfo {author} {\bibfnamefont
  {Y.}~\bibnamefont {Matsuda}},\ }\href@noop {} {\bibfield  {journal} {\bibinfo
   {journal} {Phys. Rev. X}\ }\textbf {\bibinfo {volume} {5}},\ \bibinfo
  {pages} {031022} (\bibinfo {year} {2015})}\BibitemShut {NoStop}%
\bibitem [{\citenamefont {Jiao}\ \emph {et~al.}(2017)\citenamefont {Jiao},
  \citenamefont {R\"o\ss{}ler}, \citenamefont {Koz}, \citenamefont {Schwarz},
  \citenamefont {Kasinathan}, \citenamefont {R\"o\ss{}ler},\ and\ \citenamefont
  {Wirth}}]{Jiao17}%
  \BibitemOpen
  \bibfield  {author} {\bibinfo {author} {\bibfnamefont {L.}~\bibnamefont
  {Jiao}}, \bibinfo {author} {\bibfnamefont {S.}~\bibnamefont {R\"o\ss{}ler}},
  \bibinfo {author} {\bibfnamefont {C.}~\bibnamefont {Koz}}, \bibinfo {author}
  {\bibfnamefont {U.}~\bibnamefont {Schwarz}}, \bibinfo {author} {\bibfnamefont
  {D.}~\bibnamefont {Kasinathan}}, \bibinfo {author} {\bibfnamefont {U.~K.}\
  \bibnamefont {R\"o\ss{}ler}}, \ and\ \bibinfo {author} {\bibfnamefont
  {S.}~\bibnamefont {Wirth}},\ }\href@noop {} {\bibfield  {journal} {\bibinfo
  {journal} {Phys. Rev. B}\ }\textbf {\bibinfo {volume} {96}},\ \bibinfo
  {pages} {094504} (\bibinfo {year} {2017})}\BibitemShut {NoStop}%
\bibitem [{\citenamefont {Schneider}\ and\ \citenamefont
  {Singer}(2000)}]{Schneider}%
  \BibitemOpen
  \bibfield  {author} {\bibinfo {author} {\bibfnamefont {T.}~\bibnamefont
  {Schneider}}\ and\ \bibinfo {author} {\bibfnamefont {J.~M.}\ \bibnamefont
  {Singer}},\ }\href@noop {} {\emph {\bibinfo {title} {Phase Transition
  Approach to High Temperature Superconductivity}}}\ (\bibinfo  {publisher}
  {PUBLISHED BY IMPERIAL COLLEGE PRESS AND DISTRIBUTED BY WORLD SCIENTIFIC
  PUBLISHING CO.},\ \bibinfo {year} {2000})\BibitemShut {NoStop}%
\bibitem [{\citenamefont {Chubukov}\ \emph {et~al.}(2016)\citenamefont
  {Chubukov}, \citenamefont {Eremin},\ and\ \citenamefont
  {Efremov}}]{Chubukov16}%
  \BibitemOpen
  \bibfield  {author} {\bibinfo {author} {\bibfnamefont {A.~V.}\ \bibnamefont
  {Chubukov}}, \bibinfo {author} {\bibfnamefont {I.}~\bibnamefont {Eremin}}, \
  and\ \bibinfo {author} {\bibfnamefont {D.~V.}\ \bibnamefont {Efremov}},\
  }\href@noop {} {\bibfield  {journal} {\bibinfo  {journal} {Phys. Rev. B}\
  }\textbf {\bibinfo {volume} {93}},\ \bibinfo {pages} {174516} (\bibinfo
  {year} {2016})}\BibitemShut {NoStop}%
\bibitem [{\citenamefont {Chen}\ \emph {et~al.}(2005)\citenamefont {Chen},
  \citenamefont {Stajic}, \citenamefont {Tan},\ and\ \citenamefont
  {Levin}}]{Chen05}%
  \BibitemOpen
  \bibfield  {author} {\bibinfo {author} {\bibfnamefont {Q.}~\bibnamefont
  {Chen}}, \bibinfo {author} {\bibfnamefont {J.}~\bibnamefont {Stajic}},
  \bibinfo {author} {\bibfnamefont {S.}~\bibnamefont {Tan}}, \ and\ \bibinfo
  {author} {\bibfnamefont {K.}~\bibnamefont {Levin}},\ }\href@noop {}
  {\bibfield  {journal} {\bibinfo  {journal} {Physics Reports}\ }\textbf
  {\bibinfo {volume} {412}},\ \bibinfo {pages} {1} (\bibinfo {year}
  {2005})}\BibitemShut {NoStop}%
\bibitem [{\citenamefont {Farrant}\ and\ \citenamefont {E.}(1975)}]{Farrant75}%
  \BibitemOpen
  \bibfield  {author} {\bibinfo {author} {\bibfnamefont {S.~P.}\ \bibnamefont
  {Farrant}}\ and\ \bibinfo {author} {\bibfnamefont {G.~C.}\ \bibnamefont
  {E.}},\ }\href@noop {} {\bibfield  {journal} {\bibinfo  {journal} {Phys. Rev.
  Lett.}\ }\textbf {\bibinfo {volume} {34}},\ \bibinfo {pages} {943} (\bibinfo
  {year} {1975})}\BibitemShut {NoStop}%
\bibitem [{\citenamefont {Lortz}\ \emph
  {et~al.}(2003{\natexlab{b}})\citenamefont {Lortz}, \citenamefont {Meingast},
  \citenamefont {Rykov},\ and\ \citenamefont {Tajima}}]{Lortz03}%
  \BibitemOpen
  \bibfield  {author} {\bibinfo {author} {\bibfnamefont {R.}~\bibnamefont
  {Lortz}}, \bibinfo {author} {\bibfnamefont {C.}~\bibnamefont {Meingast}},
  \bibinfo {author} {\bibfnamefont {A.~I.}\ \bibnamefont {Rykov}}, \ and\
  \bibinfo {author} {\bibfnamefont {S.}~\bibnamefont {Tajima}},\ }\href@noop {}
  {\bibfield  {journal} {\bibinfo  {journal} {Phys. Rev. Lett.}\ }\textbf
  {\bibinfo {volume} {91}},\ \bibinfo {pages} {207001} (\bibinfo {year}
  {2003}{\natexlab{b}})}\BibitemShut {NoStop}%
\bibitem [{\citenamefont {Roulin}\ \emph
  {et~al.}(1998{\natexlab{b}})\citenamefont {Roulin}, \citenamefont {Junod},
  \citenamefont {Erb},\ and\ \citenamefont {Walker}}]{Roulin98-2}%
  \BibitemOpen
  \bibfield  {author} {\bibinfo {author} {\bibfnamefont {M.}~\bibnamefont
  {Roulin}}, \bibinfo {author} {\bibfnamefont {A.}~\bibnamefont {Junod}},
  \bibinfo {author} {\bibfnamefont {A.}~\bibnamefont {Erb}}, \ and\ \bibinfo
  {author} {\bibfnamefont {E.}~\bibnamefont {Walker}},\ }\href@noop {}
  {\bibfield  {journal} {\bibinfo  {journal} {Phys. Rev. Lett.}\ }\textbf
  {\bibinfo {volume} {80}},\ \bibinfo {pages} {1722} (\bibinfo {year}
  {1998}{\natexlab{b}})}\BibitemShut {NoStop}%
\bibitem [{\citenamefont {Lortz}\ and\ \citenamefont
  {Meingast}(2002)}]{Lortz02}%
  \BibitemOpen
  \bibfield  {author} {\bibinfo {author} {\bibfnamefont {R.}~\bibnamefont
  {Lortz}}\ and\ \bibinfo {author} {\bibfnamefont {C.}~\bibnamefont
  {Meingast}},\ }\href@noop {} {\bibfield  {journal} {\bibinfo  {journal} {J.
  Non-Crystalline Solids}\ }\textbf {\bibinfo {volume} {307}},\ \bibinfo
  {pages} {452} (\bibinfo {year} {2002})}\BibitemShut {NoStop}%
\bibitem [{\citenamefont {Pippard}(1966)}]{Pippard}%
  \BibitemOpen
  \bibfield  {author} {\bibinfo {author} {\bibfnamefont {A.~B.}\ \bibnamefont
  {Pippard}},\ }\href@noop {} {\emph {\bibinfo {title} {Elements of classical
  thermodynamics for advanced students of physics}}}\ (\bibinfo  {publisher}
  {Cambridge Univ. Pr.},\ \bibinfo {address} {Cambridge [u.a.]},\ \bibinfo
  {year} {1966})\BibitemShut {NoStop}%
\bibitem [{\citenamefont {Pasler}\ \emph {et~al.}(1998)\citenamefont {Pasler},
  \citenamefont {Schweiss}, \citenamefont {Meingast}, \citenamefont {Obst},
  \citenamefont {W\"uhl}, \citenamefont {Rykov},\ and\ \citenamefont
  {Tajima}}]{Pasler98}%
  \BibitemOpen
  \bibfield  {author} {\bibinfo {author} {\bibfnamefont {V.}~\bibnamefont
  {Pasler}}, \bibinfo {author} {\bibfnamefont {P.}~\bibnamefont {Schweiss}},
  \bibinfo {author} {\bibfnamefont {C.}~\bibnamefont {Meingast}}, \bibinfo
  {author} {\bibfnamefont {B.}~\bibnamefont {Obst}}, \bibinfo {author}
  {\bibfnamefont {H.}~\bibnamefont {W\"uhl}}, \bibinfo {author} {\bibfnamefont
  {A.~I.}\ \bibnamefont {Rykov}}, \ and\ \bibinfo {author} {\bibfnamefont
  {S.}~\bibnamefont {Tajima}},\ }\href@noop {} {\bibfield  {journal} {\bibinfo
  {journal} {Phys. Rev. Lett.}\ }\textbf {\bibinfo {volume} {81}},\ \bibinfo
  {pages} {1094} (\bibinfo {year} {1998})}\BibitemShut {NoStop}%
\bibitem [{\citenamefont {Meingast}\ \emph {et~al.}(2001)\citenamefont
  {Meingast}, \citenamefont {Pasler}, \citenamefont {Nagel}, \citenamefont
  {Rykov}, \citenamefont {Tajima},\ and\ \citenamefont {Olsson}}]{Meingast01}%
  \BibitemOpen
  \bibfield  {author} {\bibinfo {author} {\bibfnamefont {C.}~\bibnamefont
  {Meingast}}, \bibinfo {author} {\bibfnamefont {V.}~\bibnamefont {Pasler}},
  \bibinfo {author} {\bibfnamefont {P.}~\bibnamefont {Nagel}}, \bibinfo
  {author} {\bibfnamefont {A.}~\bibnamefont {Rykov}}, \bibinfo {author}
  {\bibfnamefont {S.}~\bibnamefont {Tajima}}, \ and\ \bibinfo {author}
  {\bibfnamefont {P.}~\bibnamefont {Olsson}},\ }\href@noop {} {\bibfield
  {journal} {\bibinfo  {journal} {Phys. Rev. Lett.}\ }\textbf {\bibinfo
  {volume} {86}},\ \bibinfo {pages} {1606} (\bibinfo {year}
  {2001})}\BibitemShut {NoStop}%
\bibitem [{\citenamefont {Inderhees}\ \emph {et~al.}(1991)\citenamefont
  {Inderhees}, \citenamefont {Salamon}, \citenamefont {Rice},\ and\
  \citenamefont {Ginsberg}}]{Inderhees91}%
  \BibitemOpen
  \bibfield  {author} {\bibinfo {author} {\bibfnamefont {S.~E.}\ \bibnamefont
  {Inderhees}}, \bibinfo {author} {\bibfnamefont {M.~B.}\ \bibnamefont
  {Salamon}}, \bibinfo {author} {\bibfnamefont {J.~P.}\ \bibnamefont {Rice}}, \
  and\ \bibinfo {author} {\bibfnamefont {D.~M.}\ \bibnamefont {Ginsberg}},\
  }\href@noop {} {\bibfield  {journal} {\bibinfo  {journal} {Phys. Rev. Lett.}\
  }\textbf {\bibinfo {volume} {66}},\ \bibinfo {pages} {232} (\bibinfo {year}
  {1991})}\BibitemShut {NoStop}%
\bibitem [{\citenamefont {Schnelle}\ \emph {et~al.}(1993)\citenamefont
  {Schnelle}, \citenamefont {Ernst},\ and\ \citenamefont
  {Wohlleben}}]{Schnelle93}%
  \BibitemOpen
  \bibfield  {author} {\bibinfo {author} {\bibfnamefont {W.}~\bibnamefont
  {Schnelle}}, \bibinfo {author} {\bibfnamefont {P.}~\bibnamefont {Ernst}}, \
  and\ \bibinfo {author} {\bibfnamefont {D.}~\bibnamefont {Wohlleben}},\
  }\href@noop {} {\bibfield  {journal} {\bibinfo  {journal} {Ann. Phys.}\
  }\textbf {\bibinfo {volume} {2}},\ \bibinfo {pages} {109} (\bibinfo {year}
  {1993})}\BibitemShut {NoStop}%
\bibitem [{\citenamefont {Overend}\ \emph {et~al.}(1994)\citenamefont
  {Overend}, \citenamefont {Howson},\ and\ \citenamefont {Lawrie}}]{Overend94}%
  \BibitemOpen
  \bibfield  {author} {\bibinfo {author} {\bibfnamefont {N.}~\bibnamefont
  {Overend}}, \bibinfo {author} {\bibfnamefont {M.~A.}\ \bibnamefont {Howson}},
  \ and\ \bibinfo {author} {\bibfnamefont {I.~D.}\ \bibnamefont {Lawrie}},\
  }\href@noop {} {\bibfield  {journal} {\bibinfo  {journal} {Phys. Rev. Lett.}\
  }\textbf {\bibinfo {volume} {72}},\ \bibinfo {pages} {3238} (\bibinfo {year}
  {1994})}\BibitemShut {NoStop}%
\bibitem [{\citenamefont {Breit}\ \emph {et~al.}(1995)\citenamefont {Breit},
  \citenamefont {Schweiss}, \citenamefont {Hauff}, \citenamefont {W\"uhl},
  \citenamefont {Claus}, \citenamefont {Rietschel}, \citenamefont {Erb},\ and\
  \citenamefont {M\"uller-Vogt}}]{Breit95}%
  \BibitemOpen
  \bibfield  {author} {\bibinfo {author} {\bibfnamefont {V.}~\bibnamefont
  {Breit}}, \bibinfo {author} {\bibfnamefont {P.}~\bibnamefont {Schweiss}},
  \bibinfo {author} {\bibfnamefont {R.}~\bibnamefont {Hauff}}, \bibinfo
  {author} {\bibfnamefont {H.}~\bibnamefont {W\"uhl}}, \bibinfo {author}
  {\bibfnamefont {H.}~\bibnamefont {Claus}}, \bibinfo {author} {\bibfnamefont
  {H.}~\bibnamefont {Rietschel}}, \bibinfo {author} {\bibfnamefont
  {A.}~\bibnamefont {Erb}}, \ and\ \bibinfo {author} {\bibfnamefont
  {G.}~\bibnamefont {M\"uller-Vogt}},\ }\href@noop {} {\bibfield  {journal}
  {\bibinfo  {journal} {Phys. Rev. B}\ }\textbf {\bibinfo {volume} {52}},\
  \bibinfo {pages} {R15727} (\bibinfo {year} {1995})}\BibitemShut {NoStop}%
\bibitem [{\citenamefont {Junod}\ \emph {et~al.}(1994)\citenamefont {Junod},
  \citenamefont {Wang}, \citenamefont {Janod}, \citenamefont {Triscone},\ and\
  \citenamefont {Muller}}]{Junod94}%
  \BibitemOpen
  \bibfield  {author} {\bibinfo {author} {\bibfnamefont {A.}~\bibnamefont
  {Junod}}, \bibinfo {author} {\bibfnamefont {K.-Q.}\ \bibnamefont {Wang}},
  \bibinfo {author} {\bibfnamefont {E.}~\bibnamefont {Janod}}, \bibinfo
  {author} {\bibfnamefont {G.}~\bibnamefont {Triscone}}, \ and\ \bibinfo
  {author} {\bibfnamefont {J.}~\bibnamefont {Muller}},\ }\href@noop {}
  {\bibfield  {journal} {\bibinfo  {journal} {Physica B}\ }\textbf {\bibinfo
  {volume} {194}},\ \bibinfo {pages} {1495} (\bibinfo {year}
  {1994})}\BibitemShut {NoStop}%
\bibitem [{\citenamefont {Roulin}\ \emph
  {et~al.}(1995{\natexlab{b}})\citenamefont {Roulin}, \citenamefont {Junod},\
  and\ \citenamefont {Muller}}]{Roulin95-2}%
  \BibitemOpen
  \bibfield  {author} {\bibinfo {author} {\bibfnamefont {M.}~\bibnamefont
  {Roulin}}, \bibinfo {author} {\bibfnamefont {A.}~\bibnamefont {Junod}}, \
  and\ \bibinfo {author} {\bibfnamefont {J.}~\bibnamefont {Muller}},\
  }\href@noop {} {\bibfield  {journal} {\bibinfo  {journal} {Phys. Rev. Lett.}\
  }\textbf {\bibinfo {volume} {75}},\ \bibinfo {pages} {1869} (\bibinfo {year}
  {1995}{\natexlab{b}})}\BibitemShut {NoStop}%
\bibitem [{\citenamefont {Pierson}\ \emph {et~al.}(1995)\citenamefont
  {Pierson}, \citenamefont {Buan}, \citenamefont {Zhou}, \citenamefont
  {Huang},\ and\ \citenamefont {Valls}}]{Pierson95}%
  \BibitemOpen
  \bibfield  {author} {\bibinfo {author} {\bibfnamefont {S.~W.}\ \bibnamefont
  {Pierson}}, \bibinfo {author} {\bibfnamefont {J.}~\bibnamefont {Buan}},
  \bibinfo {author} {\bibfnamefont {B.}~\bibnamefont {Zhou}}, \bibinfo {author}
  {\bibfnamefont {C.~C.}\ \bibnamefont {Huang}}, \ and\ \bibinfo {author}
  {\bibfnamefont {O.~T.}\ \bibnamefont {Valls}},\ }\href@noop {} {\bibfield
  {journal} {\bibinfo  {journal} {Phys. Rev. Lett.}\ }\textbf {\bibinfo
  {volume} {74}},\ \bibinfo {pages} {1887} (\bibinfo {year}
  {1995})}\BibitemShut {NoStop}%
\bibitem [{\citenamefont {Pierson}\ \emph {et~al.}(1996)\citenamefont
  {Pierson}, \citenamefont {Katona}, \citenamefont {Tes\ifmmode \breve{}\else
  \u{}\fi{}anovi\ifmmode~\acute{c}\else \'{c}\fi{}},\ and\ \citenamefont
  {Valls}}]{Pierson96}%
  \BibitemOpen
  \bibfield  {author} {\bibinfo {author} {\bibfnamefont {S.~W.}\ \bibnamefont
  {Pierson}}, \bibinfo {author} {\bibfnamefont {T.~M.}\ \bibnamefont {Katona}},
  \bibinfo {author} {\bibfnamefont {Z.}~\bibnamefont {Tes\ifmmode \breve{}\else
  \u{}\fi{}anovi\ifmmode~\acute{c}\else \'{c}\fi{}}}, \ and\ \bibinfo {author}
  {\bibfnamefont {O.~T.}\ \bibnamefont {Valls}},\ }\href@noop {} {\bibfield
  {journal} {\bibinfo  {journal} {Phys. Rev. B}\ }\textbf {\bibinfo {volume}
  {53}},\ \bibinfo {pages} {8638} (\bibinfo {year} {1996})}\BibitemShut
  {NoStop}%
\bibitem [{\citenamefont {Pierson}\ \emph {et~al.}(1998)\citenamefont
  {Pierson}, \citenamefont {Valls}, \citenamefont {Te\ifmmode \check{s}\else
  \v{s}\fi{}anovi\ifmmode~\acute{c}\else \'{c}\fi{}},\ and\ \citenamefont
  {Lindemann}}]{Pierson98}%
  \BibitemOpen
  \bibfield  {author} {\bibinfo {author} {\bibfnamefont {S.~W.}\ \bibnamefont
  {Pierson}}, \bibinfo {author} {\bibfnamefont {O.~T.}\ \bibnamefont {Valls}},
  \bibinfo {author} {\bibfnamefont {Z.}~\bibnamefont {Te\ifmmode \check{s}\else
  \v{s}\fi{}anovi\ifmmode~\acute{c}\else \'{c}\fi{}}}, \ and\ \bibinfo {author}
  {\bibfnamefont {M.~A.}\ \bibnamefont {Lindemann}},\ }\href@noop {} {\bibfield
   {journal} {\bibinfo  {journal} {Phys. Rev. B}\ }\textbf {\bibinfo {volume}
  {57}},\ \bibinfo {pages} {8622} (\bibinfo {year} {1998})}\BibitemShut
  {NoStop}%
\bibitem [{\citenamefont {Jeandupeux}\ \emph {et~al.}(1996)\citenamefont
  {Jeandupeux}, \citenamefont {Schilling}, \citenamefont {Ott},\ and\
  \citenamefont {van Otterlo}}]{Jeandupeux96}%
  \BibitemOpen
  \bibfield  {author} {\bibinfo {author} {\bibfnamefont {O.}~\bibnamefont
  {Jeandupeux}}, \bibinfo {author} {\bibfnamefont {A.}~\bibnamefont
  {Schilling}}, \bibinfo {author} {\bibfnamefont {H.~R.}\ \bibnamefont {Ott}},
  \ and\ \bibinfo {author} {\bibfnamefont {A.}~\bibnamefont {van Otterlo}},\
  }\href@noop {} {\bibfield  {journal} {\bibinfo  {journal} {Phys. Rev. B}\
  }\textbf {\bibinfo {volume} {53}},\ \bibinfo {pages} {12475} (\bibinfo {year}
  {1996})}\BibitemShut {NoStop}%
\bibitem [{\citenamefont {Junod}\ \emph {et~al.}(2000)\citenamefont {Junod},
  \citenamefont {Roulin}, \citenamefont {Revaz},\ and\ \citenamefont
  {Erb}}]{Junod00}%
  \BibitemOpen
  \bibfield  {author} {\bibinfo {author} {\bibfnamefont {A.}~\bibnamefont
  {Junod}}, \bibinfo {author} {\bibfnamefont {M.}~\bibnamefont {Roulin}},
  \bibinfo {author} {\bibfnamefont {B.}~\bibnamefont {Revaz}}, \ and\ \bibinfo
  {author} {\bibfnamefont {A.}~\bibnamefont {Erb}},\ }\href@noop {} {\bibfield
  {journal} {\bibinfo  {journal} {Physica B}\ }\textbf {\bibinfo {volume}
  {280}},\ \bibinfo {pages} {214} (\bibinfo {year} {2000})}\BibitemShut
  {NoStop}%
\bibitem [{\citenamefont {Larkin}\ and\ \citenamefont
  {Varlamov}(2005)}]{Larkin}%
  \BibitemOpen
  \bibfield  {author} {\bibinfo {author} {\bibfnamefont {A.}~\bibnamefont
  {Larkin}}\ and\ \bibinfo {author} {\bibfnamefont {A.~A.}\ \bibnamefont
  {Varlamov}},\ }\href@noop {} {\emph {\bibinfo {title} {Theory of fluctuations
  in superconductors}}},\ edited by\ \bibinfo {editor} {\bibfnamefont {A.~I.}\
  \bibnamefont {Larkin}},\ International series of monographs on physics ;
  127Oxford science publications\ (\bibinfo  {publisher} {Clarendon Press},\
  \bibinfo {address} {Oxford},\ \bibinfo {year} {2005})\BibitemShut {NoStop}%
\bibitem [{\citenamefont {Li}\ and\ \citenamefont
  {Rosenstein}(2001{\natexlab{a}})}]{Li01}%
  \BibitemOpen
  \bibfield  {author} {\bibinfo {author} {\bibfnamefont {D.}~\bibnamefont
  {Li}}\ and\ \bibinfo {author} {\bibfnamefont {B.}~\bibnamefont
  {Rosenstein}},\ }\href@noop {} {\bibfield  {journal} {\bibinfo  {journal}
  {Phys. Rev. B}\ }\textbf {\bibinfo {volume} {65}},\ \bibinfo {pages} {024514}
  (\bibinfo {year} {2001}{\natexlab{a}})}\BibitemShut {NoStop}%
\bibitem [{\citenamefont {Li}\ and\ \citenamefont
  {Rosenstein}(2001{\natexlab{b}})}]{Li01-2}%
  \BibitemOpen
  \bibfield  {author} {\bibinfo {author} {\bibfnamefont {D.}~\bibnamefont
  {Li}}\ and\ \bibinfo {author} {\bibfnamefont {B.}~\bibnamefont
  {Rosenstein}},\ }\href@noop {} {\bibfield  {journal} {\bibinfo  {journal}
  {Phys. Rev. Lett.}\ }\textbf {\bibinfo {volume} {86}},\ \bibinfo {pages}
  {3618} (\bibinfo {year} {2001}{\natexlab{b}})}\BibitemShut {NoStop}%
\bibitem [{\citenamefont {Li}\ and\ \citenamefont {Rosenstein}(2003)}]{Li03}%
  \BibitemOpen
  \bibfield  {author} {\bibinfo {author} {\bibfnamefont {D.}~\bibnamefont
  {Li}}\ and\ \bibinfo {author} {\bibfnamefont {B.}~\bibnamefont
  {Rosenstein}},\ }\href@noop {} {\bibfield  {journal} {\bibinfo  {journal}
  {Phys. Rev. Lett.}\ }\textbf {\bibinfo {volume} {90}},\ \bibinfo {pages}
  {167004} (\bibinfo {year} {2003})}\BibitemShut {NoStop}%
\bibitem [{\citenamefont {Li}\ and\ \citenamefont {Rosenstein}(2004)}]{Li04}%
  \BibitemOpen
  \bibfield  {author} {\bibinfo {author} {\bibfnamefont {D.}~\bibnamefont
  {Li}}\ and\ \bibinfo {author} {\bibfnamefont {B.}~\bibnamefont
  {Rosenstein}},\ }\href@noop {} {\bibfield  {journal} {\bibinfo  {journal}
  {Phys. Rev. B}\ }\textbf {\bibinfo {volume} {70}},\ \bibinfo {pages} {144521}
  (\bibinfo {year} {2004})}\BibitemShut {NoStop}%
\bibitem [{\citenamefont {Rosenstein}\ and\ \citenamefont
  {Li}(2010)}]{Rosenstein10}%
  \BibitemOpen
  \bibfield  {author} {\bibinfo {author} {\bibfnamefont {B.}~\bibnamefont
  {Rosenstein}}\ and\ \bibinfo {author} {\bibfnamefont {D.}~\bibnamefont
  {Li}},\ }\href@noop {} {\bibfield  {journal} {\bibinfo  {journal} {Rev. Mod.
  Phys.}\ }\textbf {\bibinfo {volume} {82}},\ \bibinfo {pages} {109} (\bibinfo
  {year} {2010})}\BibitemShut {NoStop}%
\bibitem [{\citenamefont {Thouless}(1975)}]{Thouless75}%
  \BibitemOpen
  \bibfield  {author} {\bibinfo {author} {\bibfnamefont {D.~J.}\ \bibnamefont
  {Thouless}},\ }\href@noop {} {\bibfield  {journal} {\bibinfo  {journal}
  {Phys. Rev. Lett.}\ }\textbf {\bibinfo {volume} {34}},\ \bibinfo {pages}
  {946} (\bibinfo {year} {1975})}\BibitemShut {NoStop}%
\bibitem [{\citenamefont {Mikitik}\ and\ \citenamefont
  {Brandt}(2003)}]{Mikitik03}%
  \BibitemOpen
  \bibfield  {author} {\bibinfo {author} {\bibfnamefont {G.~P.}\ \bibnamefont
  {Mikitik}}\ and\ \bibinfo {author} {\bibfnamefont {E.~H.}\ \bibnamefont
  {Brandt}},\ }\href@noop {} {\bibfield  {journal} {\bibinfo  {journal} {Phys.
  Rev. B}\ }\textbf {\bibinfo {volume} {68}},\ \bibinfo {pages} {054509}
  (\bibinfo {year} {2003})}\BibitemShut {NoStop}%
\bibitem [{\citenamefont {Sun}\ \emph {et~al.}(2015)\citenamefont {Sun},
  \citenamefont {Pyon}, \citenamefont {Tamegai}, \citenamefont {Kobayashi},
  \citenamefont {Watashige}, \citenamefont {Kasahara}, \citenamefont
  {Matsuda},\ and\ \citenamefont {Shibauchi}}]{Sun15}%
  \BibitemOpen
  \bibfield  {author} {\bibinfo {author} {\bibfnamefont {Y.}~\bibnamefont
  {Sun}}, \bibinfo {author} {\bibfnamefont {S.}~\bibnamefont {Pyon}}, \bibinfo
  {author} {\bibfnamefont {T.}~\bibnamefont {Tamegai}}, \bibinfo {author}
  {\bibfnamefont {R.}~\bibnamefont {Kobayashi}}, \bibinfo {author}
  {\bibfnamefont {T.}~\bibnamefont {Watashige}}, \bibinfo {author}
  {\bibfnamefont {S.}~\bibnamefont {Kasahara}}, \bibinfo {author}
  {\bibfnamefont {Y.}~\bibnamefont {Matsuda}}, \ and\ \bibinfo {author}
  {\bibfnamefont {T.}~\bibnamefont {Shibauchi}},\ }\href@noop {} {\bibfield
  {journal} {\bibinfo  {journal} {Phys. Rev. B}\ }\textbf {\bibinfo {volume}
  {92}},\ \bibinfo {pages} {144509} (\bibinfo {year} {2015})}\BibitemShut
  {NoStop}%
\bibitem [{\citenamefont {Koshelev}\ and\ \citenamefont
  {Varlamov}(2014)}]{Koshelev14}%
  \BibitemOpen
  \bibfield  {author} {\bibinfo {author} {\bibfnamefont {A.~E.}\ \bibnamefont
  {Koshelev}}\ and\ \bibinfo {author} {\bibfnamefont {A.~A.}\ \bibnamefont
  {Varlamov}},\ }\href@noop {} {\bibfield  {journal} {\bibinfo  {journal}
  {Supercond. Sci. Technol.}\ }\textbf {\bibinfo {volume} {27}},\ \bibinfo
  {pages} {124001} (\bibinfo {year} {2014})}\BibitemShut {NoStop}%
\bibitem [{\citenamefont {Adachi}\ and\ \citenamefont
  {Ikeda}(2016)}]{Adachi16}%
  \BibitemOpen
  \bibfield  {author} {\bibinfo {author} {\bibfnamefont {K.}~\bibnamefont
  {Adachi}}\ and\ \bibinfo {author} {\bibfnamefont {R.}~\bibnamefont {Ikeda}},\
  }\href@noop {} {\bibfield  {journal} {\bibinfo  {journal} {Phys. Rev. B}\
  }\textbf {\bibinfo {volume} {93}},\ \bibinfo {pages} {134503} (\bibinfo
  {year} {2016})}\BibitemShut {NoStop}%
\bibitem [{\citenamefont {Ok}\ \emph {et~al.}(2020)\citenamefont {Ok},
  \citenamefont {Kwon}, \citenamefont {Kohama}, \citenamefont {You},
  \citenamefont {Park}, \citenamefont {hye Kim}, \citenamefont {Jo},
  \citenamefont {Choi}, \citenamefont {Kindo}, \citenamefont {Kang},
  \citenamefont {Kim}, \citenamefont {Moon}, \citenamefont {Gurevich},\ and\
  \citenamefont {Kim}}]{Korean20}%
  \BibitemOpen
  \bibfield  {author} {\bibinfo {author} {\bibfnamefont {J.~M.}\ \bibnamefont
  {Ok}}, \bibinfo {author} {\bibfnamefont {C.~I.}\ \bibnamefont {Kwon}},
  \bibinfo {author} {\bibfnamefont {Y.}~\bibnamefont {Kohama}}, \bibinfo
  {author} {\bibfnamefont {J.~S.}\ \bibnamefont {You}}, \bibinfo {author}
  {\bibfnamefont {S.~K.}\ \bibnamefont {Park}}, \bibinfo {author}
  {\bibfnamefont {J.}~\bibnamefont {hye Kim}}, \bibinfo {author} {\bibfnamefont
  {Y.~J.}\ \bibnamefont {Jo}}, \bibinfo {author} {\bibfnamefont {E.~S.}\
  \bibnamefont {Choi}}, \bibinfo {author} {\bibfnamefont {K.}~\bibnamefont
  {Kindo}}, \bibinfo {author} {\bibfnamefont {W.}~\bibnamefont {Kang}},
  \bibinfo {author} {\bibfnamefont {K.~S.}\ \bibnamefont {Kim}}, \bibinfo
  {author} {\bibfnamefont {E.~G.}\ \bibnamefont {Moon}}, \bibinfo {author}
  {\bibfnamefont {A.}~\bibnamefont {Gurevich}}, \ and\ \bibinfo {author}
  {\bibfnamefont {J.~S.}\ \bibnamefont {Kim}},\ }\href@noop {} {} (\bibinfo
  {year} {2020}),\ \Eprint {http://arxiv.org/abs/2003.12351} {arXiv:2003.12351
  [cond-mat.supr-con]} \BibitemShut {NoStop}%
\bibitem [{\citenamefont {Maki}(1964)}]{Maki64}%
  \BibitemOpen
  \bibfield  {author} {\bibinfo {author} {\bibfnamefont {K.}~\bibnamefont
  {Maki}},\ }\href@noop {} {\bibfield  {journal} {\bibinfo  {journal} {Physics
  (Long Island City, N.Y.)}\ }\textbf {\bibinfo {volume} {1}},\ \bibinfo
  {pages} {127} (\bibinfo {year} {1964})}\BibitemShut {NoStop}%
\bibitem [{\citenamefont {Sarma}(1963)}]{Sarma63}%
  \BibitemOpen
  \bibfield  {author} {\bibinfo {author} {\bibfnamefont {G.}~\bibnamefont
  {Sarma}},\ }\href@noop {} {\bibfield  {journal} {\bibinfo  {journal} {J.
  Phys. Chem. Solids}\ }\textbf {\bibinfo {volume} {24}},\ \bibinfo {pages}
  {1029} (\bibinfo {year} {1963})}\BibitemShut {NoStop}%
\bibitem [{\citenamefont {Helfand}\ and\ \citenamefont
  {Werthamer}(1966)}]{Helfand66}%
  \BibitemOpen
  \bibfield  {author} {\bibinfo {author} {\bibfnamefont {E.}~\bibnamefont
  {Helfand}}\ and\ \bibinfo {author} {\bibfnamefont {N.~R.}\ \bibnamefont
  {Werthamer}},\ }\href@noop {} {\bibfield  {journal} {\bibinfo  {journal}
  {Phys. Rev.}\ }\textbf {\bibinfo {volume} {147}},\ \bibinfo {pages} {288}
  (\bibinfo {year} {1966})}\BibitemShut {NoStop}%
\bibitem [{\citenamefont {Werthamer}\ \emph {et~al.}(1966)\citenamefont
  {Werthamer}, \citenamefont {Helfand},\ and\ \citenamefont
  {Hohenberg}}]{Werthamer66}%
  \BibitemOpen
  \bibfield  {author} {\bibinfo {author} {\bibfnamefont {N.~R.}\ \bibnamefont
  {Werthamer}}, \bibinfo {author} {\bibfnamefont {E.}~\bibnamefont {Helfand}},
  \ and\ \bibinfo {author} {\bibfnamefont {P.~C.}\ \bibnamefont {Hohenberg}},\
  }\href@noop {} {\bibfield  {journal} {\bibinfo  {journal} {Phys. Rev.}\
  }\textbf {\bibinfo {volume} {147}},\ \bibinfo {pages} {295} (\bibinfo {year}
  {1966})}\BibitemShut {NoStop}%
\bibitem [{\citenamefont {Brison}\ \emph {et~al.}(1995)\citenamefont {Brison},
  \citenamefont {Keller}, \citenamefont {Vernière}, \citenamefont {Lejay},
  \citenamefont {Schmidt}, \citenamefont {Buzdin}, \citenamefont {Flouquet},
  \citenamefont {Julian},\ and\ \citenamefont {Lonzarich}}]{Brison95}%
  \BibitemOpen
  \bibfield  {author} {\bibinfo {author} {\bibfnamefont {J.}~\bibnamefont
  {Brison}}, \bibinfo {author} {\bibfnamefont {N.}~\bibnamefont {Keller}},
  \bibinfo {author} {\bibfnamefont {A.}~\bibnamefont {Vernière}}, \bibinfo
  {author} {\bibfnamefont {P.}~\bibnamefont {Lejay}}, \bibinfo {author}
  {\bibfnamefont {L.}~\bibnamefont {Schmidt}}, \bibinfo {author} {\bibfnamefont
  {A.}~\bibnamefont {Buzdin}}, \bibinfo {author} {\bibfnamefont
  {J.}~\bibnamefont {Flouquet}}, \bibinfo {author} {\bibfnamefont
  {S.}~\bibnamefont {Julian}}, \ and\ \bibinfo {author} {\bibfnamefont
  {G.}~\bibnamefont {Lonzarich}},\ }\href@noop {} {\bibfield  {journal}
  {\bibinfo  {journal} {Physica C}\ }\textbf {\bibinfo {volume} {250}},\
  \bibinfo {pages} {128} (\bibinfo {year} {1995})}\BibitemShut {NoStop}%
\bibitem [{\citenamefont {Houghton}\ \emph {et~al.}(1989)\citenamefont
  {Houghton}, \citenamefont {Pelcovits},\ and\ \citenamefont
  {Sudb\o{}}}]{Houghton89}%
  \BibitemOpen
  \bibfield  {author} {\bibinfo {author} {\bibfnamefont {A.}~\bibnamefont
  {Houghton}}, \bibinfo {author} {\bibfnamefont {R.~A.}\ \bibnamefont
  {Pelcovits}}, \ and\ \bibinfo {author} {\bibfnamefont {A.}~\bibnamefont
  {Sudb\o{}}},\ }\href@noop {} {\bibfield  {journal} {\bibinfo  {journal}
  {Phys. Rev. B}\ }\textbf {\bibinfo {volume} {40}},\ \bibinfo {pages} {6763}
  (\bibinfo {year} {1989})}\BibitemShut {NoStop}%
\bibitem [{\citenamefont {I.M.~Babich}\ and\ \citenamefont
  {Mikitik}(1994)}]{Babich94}%
  \BibitemOpen
  \bibfield  {author} {\bibinfo {author} {\bibfnamefont {Y.~S.}\ \bibnamefont
  {I.M.~Babich}}\ and\ \bibinfo {author} {\bibfnamefont {G.}~\bibnamefont
  {Mikitik}},\ }\href@noop {} {\bibfield  {journal} {\bibinfo  {journal} {Low.
  Temp. Phys.}\ }\textbf {\bibinfo {volume} {20}},\ \bibinfo {pages} {221}
  (\bibinfo {year} {1994})}\BibitemShut {NoStop}%
\bibitem [{\citenamefont {Gruenberg}\ and\ \citenamefont
  {Gunther}(1966)}]{Gruenberg66}%
  \BibitemOpen
  \bibfield  {author} {\bibinfo {author} {\bibfnamefont {L.~W.}\ \bibnamefont
  {Gruenberg}}\ and\ \bibinfo {author} {\bibfnamefont {L.}~\bibnamefont
  {Gunther}},\ }\href@noop {} {\bibfield  {journal} {\bibinfo  {journal} {Phys.
  Rev. Lett.}\ }\textbf {\bibinfo {volume} {16}},\ \bibinfo {pages} {996}
  (\bibinfo {year} {1966})}\BibitemShut {NoStop}%
\bibitem [{\citenamefont {Houzet}\ and\ \citenamefont
  {Buzdin}(2001)}]{Houzet01}%
  \BibitemOpen
  \bibfield  {author} {\bibinfo {author} {\bibfnamefont {M.}~\bibnamefont
  {Houzet}}\ and\ \bibinfo {author} {\bibfnamefont {A.}~\bibnamefont
  {Buzdin}},\ }\href@noop {} {\bibfield  {journal} {\bibinfo  {journal} {Phys.
  Rev. B}\ }\textbf {\bibinfo {volume} {63}},\ \bibinfo {pages} {184521}
  (\bibinfo {year} {2001})}\BibitemShut {NoStop}%
\bibitem [{\citenamefont {Houzet}\ and\ \citenamefont
  {Mineev}(2006)}]{Houzet06}%
  \BibitemOpen
  \bibfield  {author} {\bibinfo {author} {\bibfnamefont {M.}~\bibnamefont
  {Houzet}}\ and\ \bibinfo {author} {\bibfnamefont {V.~P.}\ \bibnamefont
  {Mineev}},\ }\href@noop {} {\bibfield  {journal} {\bibinfo  {journal} {Phys.
  Rev. B}\ }\textbf {\bibinfo {volume} {74}},\ \bibinfo {pages} {144522}
  (\bibinfo {year} {2006})}\BibitemShut {NoStop}%
\bibitem [{\citenamefont {Houzet}\ and\ \citenamefont
  {Mineev}(2007)}]{Houzet07}%
  \BibitemOpen
  \bibfield  {author} {\bibinfo {author} {\bibfnamefont {M.}~\bibnamefont
  {Houzet}}\ and\ \bibinfo {author} {\bibfnamefont {V.~P.}\ \bibnamefont
  {Mineev}},\ }\href@noop {} {\bibfield  {journal} {\bibinfo  {journal} {Phys.
  Rev. B}\ }\textbf {\bibinfo {volume} {76}},\ \bibinfo {pages} {224508}
  (\bibinfo {year} {2007})}\BibitemShut {NoStop}%
\bibitem [{\citenamefont {A.I.~Buzdin}(1996)}]{Buzdin96}%
  \BibitemOpen
  \bibfield  {author} {\bibinfo {author} {\bibfnamefont {J.~B.}\ \bibnamefont
  {A.I.~Buzdin}},\ }\href@noop {} {\bibfield  {journal} {\bibinfo  {journal}
  {Phys. Lett. A}\ }\textbf {\bibinfo {volume} {218}},\ \bibinfo {pages} {359}
  (\bibinfo {year} {1996})}\BibitemShut {NoStop}%
\end{thebibliography}%

\end{document}